\documentclass[%
 reprint,
showpacs,preprintnumbers,
nofootinbib,
nobibnotes,
 amsmath,amssymb,
 aps,
prd,
floatfix,
]{revtex4-1}

\usepackage[colorlinks=false]{hyperref}
\usepackage{graphicx}
\usepackage{dcolumn}
\usepackage{bm}

\usepackage{graphicx,color,xcolor}
\usepackage{enumitem}

\DeclareMathOperator{\M}{\mathcal{M}}
\DeclareMathOperator{\E}{\mathcal{E}}

\DeclareMathOperator{\df}{\mathrm{d}\!}

\DeclareMathOperator{\K}{\mathcal{K}}

\DeclareMathOperator{\n}{\text{n}}
\DeclareMathOperator{\p}{\text{p}}

\def\eul{{\mathfrak{n}}}      

\begin{document}


\title[Realistic superfluid neutron stars in general
relativity]{Numerical models for stationary superfluid neutron stars in general relativity with realistic equations of state}

\author{Aur\'elien Sourie }\email{aurelien.sourie@obspm.fr}
\author{Micaela Oertel}\email{micaela.oertel@obspm.fr}
\author{J\'er\^ome Novak}\email{jerome.novak@obspm.fr}

\affiliation{LUTH, Observatoire de Paris, PSL Research University, CNRS,
Universit\'e Paris Diderot, Sorbonne Paris Cit\'e, 5 place Jules Janssen, 92195 Meudon }%

\date{\today}

\begin{abstract}

  We present a numerical model for uniformly rotating superfluid
  neutron stars, for the first time with realistic microphysics
  including entrainment, in a fully general relativistic framework. We
  compute stationary and axisymmetric configurations of neutron stars
  composed of two fluids, namely superfluid neutrons and charged
  particles (protons and electrons), rotating with different rates
  around a common axis. Both fluids are coupled by entrainment, a
  non-dissipative interaction which in case of a non-vanishing
  relative velocity between the fluids, causes the fluid momenta being
  not aligned with the respective fluid velocities. We extend the
  formalism by Comer and Joynt~\cite{comer2003relativistic} in order
  to calculate the equation of state (EoS) and entrainment parameters
  for an arbitrary relative velocity as far as superfluidity is
  maintained. The resulting entrainment matrix fulfills all necessary
  sum rules and in the limit of small relative velocity our results
  agree with Fermi liquid theory ones, derived to lowest order in the
  velocity. This formalism is applied to two new nuclear equations of
  state which are implemented in the numerical model. We are able to
  obtain precise equilibrium configurations. Resulting density
  profiles and moments of inertia are discussed employing both EoSs,
  showing the impact of entrainment and the dependence on the EoS.


\end{abstract}

\pacs{97.60.Jd,  
26.60.-c, 	
26.60.Dd, 	
04.25.D-, 	
04.40.Dg  
}

\maketitle



\section{ \label{sec:intro} Introduction}

Spanning over fifteen orders of magnitude in density, the composition
of a neutron star is quite complex
\cite{haensel2007neutron}. Migdal~\cite{migdal1959superfluidity} first
suggested the possibility that superfluidity could appear in neutron
star matter at sufficiently low temperature, through the formation of
neutron Cooper pairs. From detailed microscopic calculations
(e.g. \citep{dean2003pairing}), the superfluid critical temperature
has been estimated to be of the order of $\sim10^{9}-10^{10}$ K. As a
neutron star typically drops below this temperature within a few years
after its birth \citep{gnedin2001thermal}, neutrons are supposed to
form a superfluid in the core and in the inner crust of the
star. Protons are likely to form a superconducting fluid in the core,
too.

The presence of superfluid matter in the interior of neutron stars
is strongly supported by the qualitative success of superfluid models
\citep{anderson1975pulsar, alpar1984vortex, haskell2012modelling} to
explain the observed features of pulsar glitches and, especially, the
very long relaxation time scales \citep{wong2001observations,
  espinoza2011study} (see \cite{haskell2015models} for a review on
models for pulsar glitches). The recent direct observations of the
fast cooling of the young neutron star in the Cassiopeia A supernova
remnant \citep{heinke2010direct, shternin2011cooling} also provide
serious evidence for nucleon superfluidity in the core of neutron
stars \citep{page2011rapid, shternin2011cooling}.  Moreover, the
quasi-periodic oscillations detected in the X-ray flux of giant flares
from some soft gamma-ray repeaters (see \cite{strohmayer20062004}, for
instance) have been interpreted as the signature of superfluid
magneto-elastic oscillations \cite{gabler2013imprints}, bringing thus
a new, albeit less convincing, observational support for
superfluidity.

Due to superfluidity, the matter inside the star has to be described
as a mixture of several species with different dynamics. A first fluid
is supposed to be made of superfluid neutrons in the crust and the
outer core, which can ``freely" flow through the other components. On
the other hand, protons, nuclei in the crust, electrons and possibly
muons are locked together on very short time scales by short-range
electromagnetic interactions, forming a fluid of charged particles,
called here simply ``protons''. Being coupled to the magnetosphere
through magnetic effects, this fluid is rotating at the observed
angular velocity of the star. The above statements correspond to the
so-called \textit{two-fluid model} for the interior of neutron stars
\citep{baym1969superfluidity}. Although rotating around a common axis
with (possibly) different angular velocities, neutron and proton
fluids do not strictly flow independently, but are rather coupled
through \textit{entrainment}. While in the core this non-dissipative
phenomenon arises from the strong interactions between neutrons and
protons \citep{alpar1984rapid, chamel2008two}, entrainment in the
inner crust comes from Bragg scattering of dripped neutrons by nuclei
\citep{chamel2005band, chamel2012neutron}, leading to much more
important effects. Entrainment is an important ingredient in the
understanding of oscillations of superfluid neutron stars (acting on
both frequency and damping rate \citep{andersson2002oscillations,
  haskell2009r}) and pulsar glitches \citep{andersson2012pulsar,
  chamel2013crustal}.

Based on the elegant formalism developed by Carter and coworkers
\citep{carter1989covariant, carter1998relativistic,
  langlois1998differential}, a lot of progress has been made in the
past few years to obtain realistic equilibrium configurations of
two-fluid neutron stars, in a fully relativistic framework. These
models are not only interesting for the study of stationary properties
of superfluid neutron stars, but can also be useful as unperturbed
initial states for dynamical simulations of neutron star oscillations
or collapse to black holes. For the first time, Andersson and
Comer~\cite{andersson2001slowly} computed stationary configurations in
the slow rotation approximation, using an analytic equation of state
(EoS). This work was then extended by Comer and
Joynt~\cite{comer2003relativistic, comer2004slowly} who considered a
simplified nuclear EoS model, including entrainment effects. More
recently, several improvements were made to get more realistic EoSs
\citep{gusakov2009relativistic, kheto2014isospin, kheto2015slowly},
including in particular the correct interaction for isospin asymmetric
neutron star matter. Meanwhile, Prix \textit{et
  al.}~\cite{prix2005relativistic} have built the first complete
numerical solutions of stationary rotating superfluid neutron stars,
for any rotation rates. Going beyond the slow rotation approximation
is particularly interesting as several pulsars are observed to be
rapidly rotating, with angular frequencies up to 716 Hz
\citep{hessels2006radio}, corresponding to a surface velocity at the
equator of the order of $\sim c/6$ (assuming $R\sim 12$ km). Yet, only
polytropic EoSs were considered in Prix \textit{et
  al.}~\citep{prix2005relativistic}, for better numerical convergence.

Here, we present realistic stationary and axisymmetric configurations
of rotating superfluid neutron stars, in a full general relativistic
framework, extending the work by Prix \textit{et
  al.}~\cite{prix2005relativistic} by implementing two new realistic
EoSs. These are density-dependent relativistic mean-field models
\citep{typel1999relativistic, avancini2009nuclear}, that we adapted to
a system of two fluids coupled by entrainment. Our derivation of the
EoS with entrainment follows the spirit of
\cite{comer2003relativistic}, with the difference that we choose the
neutron rest frame instead of the neutron zero-momentum frame for our
calculations. This allows to compute in a very convenient way the EoS
to any order in the spatial velocity of the proton current, {\it i.e.}
the relative velocity between the two fluids. In contrast to the
results of \cite{comer2003relativistic,kheto2014isospin}, the
resulting entrainment matrix fulfills all relations required by
spacetime symmetries and the slow velocity approximation is in
agreement with the result of ~\cite{gusakov2009relativistic} derived
from relativistic Fermi liquid theory to lowest order in the relative
velocity.

The paper is organized as follows. In Section II, we present the major
assumptions employed in our model and we recall the main features of
two-fluid hydrodynamics. In Section III, we explain our formalism to
calculate the EoS with entrainment and describe the two new EoSs we
use to compute equilibrium configurations. These configurations are
then presented in Section IV. Finally, a discussion of this work is
given in Section V.  Throughout this paper, gravitational units, $ G =
c = \hbar = 1$, are adopted. The signature of the spacetime metric is
given by $(-, +, +, +)$. Greek indices $\alpha$, $\beta$, \dots, $\mu,
\nu$, \dots\ are used to refer to space and time components
$\{0,1,2,3\}$ of a tensor, whereas Latin indices $i$, $j$, \dots\
stand for spatial terms $\{1,2,3\}$ only. Einstein summation
convention is used on repeated indices, except when the capital
letters $X$ and $Y$ referring to the two fluids are employed. Isospin
vectors are denoted by an arrow: e.g. $\vec{\delta}$.


\section[]{Two-fluid model}
\label{two-fluid}

\subsection[]{Global  framework }
\label{assumptions}

As a simplified composition, we only consider a uniform mixture of
neutrons, protons and electrons. Such a composition is likely to be
found in the outer core of neutron stars, corresponding to densities
ranging from $\sim \rho_0/2$ to $\sim 2-3 \rho_0$, where $\rho_0\simeq
2.8 \times 10^4$ g.cm$^{-3}$ denotes the saturation density of
infinite symmetric nuclear matter. Here, we simply assume that it
remains the same at all densities. Note that muons could be included
straightforwardly in our model, but are not expected to strongly
affect the global properties of the star. The composition of the inner
core being still poorly known, we do not consider the possible
appearance of any additional particle. Furthermore, the presence of
the solid crust is also neglected. Even though a relativistic
description unifying the core and the inner crust within a two-fluid
context exists \citep{carter2005entrainment, carter2006entrainment,
  carter2006relativistic}, computing realistic configurations would
require a suited EoS, which is beyond the scope of the present work.

Even soon after their birth, typical temperatures of neutron stars are
much smaller than the Fermi energy of the interior, which can be
assumed to be greater than $\sim 60$ MeV (\textit{i.e.} $T \sim 7
\times 10^{11}$ K) for a density exceeding the nuclear one
(e.g. \citep{friedman2013rotating}), indicating that finite
temperature effects can be neglected on the EoS. In this sense,
neutron stars are cold and can be reasonably well described by a
zero-temperature EoS. Assuming null temperature, all the neutrons will
therefore be in a superfluid state. We assume in addition that the
temperature lies well below the critical temperature of (neutron)
superfluidity, such that temperature effects on entrainment can be
neglected, too, see \cite{Gusakovtemperature} for a discussion.

In our model, the magnetic field of the star is only considered by
requiring that the electromagnetically charged particles are
comoving\footnote{Strictly speaking, this assumption is only valid on
  time scales larger than a few seconds \cite{easson1979postglitch}.
  This question has been recently discussed by Glampedakis and
  Lasky~\cite{glampedakis2015persistent}.} (see
Sec.~\ref{sec:intro}). Consequently, our system shall be described by
two fluids: superfluid neutrons, labeled by ``n'', and ``normal''
matter in form of protons and electrons, labeled by ``p''. The effect
of magnetic field on the EoS is anyway expected to be negligible and
its influence on the global structure very small, except maybe for
some extreme magnetars \citep{chatterjee2015consistent}. Including the
magnetic field in our model, which would require a better
understanding of proton superconductivity, is thus left for future
work.

In our study of equilibrium configurations, we neglect any kind of
dissipating mechanisms, which would prevent the star from being in a
stationary state. Consequently, we do not consider any departure from
pressure isotropy due to crustal and magnetic stresses nor heat flow
(see above). Possible transfer of matter between the fluids, known as
transfusion process (see \citep{langlois1998differential} and
Sec.~\ref{section_structure}), is not taken into account and we assume
the viscosity of charged particles to be very small, so that we can
reasonably neglect it. Moreover, being superfluid, the vorticity of
the neutrons is confined to vortex lines, whose interactions with the
surrounding medium leads to dissipative processes, such as pinning or
mutual friction forces, which are not considered here. We thus make
the assumption that the stationary configurations of a superfluid
neutron star can accurately be described by two \textit{perfect}
fluids \citep{friedman2013rotating}. Doing so, we do not take the
presence of the superfluid vortices into account in our model. This
assumption only makes sense on scales much larger than the intervortex
spacing, typically on a few centimeters, on which the presence of this
array of vortices mimics rigid-body rotation.
 
We consider a general relativistic framework, following Bonazzola
\textit{et al.}~\cite{bonazzola1993axisymmetric} and we assume the
neutron star spacetime $(\mathcal{M},\ g_{\mu\nu})$ to be
\textit{stationary}, \textit{axisymmetric} and \textit{asymptotically
  flat}. The two symmetries, stationarity and axisymmetry, are
respectively associated with the Killing vector fields $\xi^\mu$,
timelike at spatial infinity, and $\chi^\mu$, spacelike everywhere and
vanishing on the rotation axis of the star. We choose spherical-type
coordinate system $\left(x^0 = t, x^1=r, x^2=\theta,
  x^3=\varphi\right)$, such that $\xi^\mu =\partial_t^{\ \mu}$ and
$\chi^\mu = \partial_\varphi^{\ \mu}$. Furthermore, we also assume
that the spacetime is \textit{circular}. This implies that the
energy-momentum tensor $T^{\mu\nu}$ has to verify conditions given by
the generalized Papapetrou theorem
\citep{bonazzola1993axisymmetric}. As long as the interior of neutron
stars is described by perfect fluids, these conditions lead to
consider only purely circular motion around the rotation axis, with
angular velocities $\Omega_{\n}$ and $\Omega_{\p}$. Thus, no
convection is allowed. Choosing quasi-isotropic coordinates, the line
element of a rotating neutron star at equilibrium, under the previous
assumptions, reads:
\begin{equation}
\label{metrique}
  \begin{array}{rcl}
    ds^2 &=& g_{\mu\nu}\df x^{\mu}\df x^{\nu}\\[+3pt]
    &=& -N^2\df t^2 + A^2(\df r^2+r^2\df\theta^2) \\[+3pt]
    &\ & + B^2r^2\sin^2\theta(\df\varphi -\omega \df t)^2
  \end{array}
\end{equation}
where $g_{\mu\nu}$ denotes the spacetime metric whose components $N$,
$A$, $B$ and $\omega$ are four functions depending only on $r$ and
$\theta$.

Finally, we assume both fluids to be rigidly rotating. Although
neutron stars are likely to present differential rotation at birth,
several mechanisms are said to enforce rigid rotation: magnetic
braking suppresses differential rotation on a time scale of tens of
seconds \citep{shapiro2000differential}; viscous dissipation, caused
by kinematic shear viscosity, enforces uniform rotation on a much
longer time scale of the order of years \citep{flowers1976transport};
turbulence mixing may also suppress any amount of differential
rotation within a few days \citep{hegyi1977upper}. So, it seems
reasonable to consider $\Omega_{\p}$ to be uniform. Nevertheless, one
must notice that some amount of differential rotation is likely to be
present when dynamical time scales are shorter than typical damping
time scales, during glitches or oscillations for instance. For the
sake of simplicity, we also consider that $\Omega_{\n}$ is uniform,
although the damping mechanisms presented above do not play any role
in a superfluid.

\subsection[]{Two-fluid hydrodynamics}
\label{hydro}

Our model is based on the covariant formalism developed by Carter and
collaborators \citep{carter1989covariant, carter1998relativistic,
  langlois1998differential}, who described a system made of two
perfect fluids coupled by entrainment in a general relativistic
framework. Here, we recall briefly the main features of this model;
more details can be found in Prix \textit{et
  al.}~\cite{prix2005relativistic}.

Following this approach, the two fluids are described, at macroscopic
scales, with mean 4-velocity fields $u_{\n}^{\ \mu}$ and $u_{\p}^{\
  \mu}$ or equivalently with average particle 4-currents $n_{\n}^{\
  \mu}$ and $n_{\p}^{\ \mu}$. Since dissipative effects are neglected,
this system can be studied in terms of a variational principle based
on a Lagrangian density $\Lambda$ which depends on the two quantities
$n_{\n}^\mu$ and $n_{\p}^\mu$. $\Lambda$ is commonly referred to as
the \textit{master function}, because it contains all the information
relative to the local thermodynamic state of the system. From
covariance requirement, $\Lambda$ only depends on the three scalars
that can be formed from the particle 4-currents
\begin{equation}
\label{scalaire}
n_{\n}^2 = -  n_{\n}^{\ \mu} n_{\n\,\mu}, \ \ n_{\p}^2 =
-n_{\p}^{\ \mu} n_{\p\,\mu} \ \ \text{and} \ \ x^2 =
-n_{\n}^{\ \mu} n_{\p\,\mu}~.  
\end{equation}
Thus, the Lagrangian density can be written as 
\begin{equation}
\label{covariance}
\Lambda(n_{\n}^{\ \mu} , n_{\p}^{\ \mu}) = - \E ( n_{\n}^2, n_{\p}^2, x^2),
\end{equation} 
where $\E$ refers to the total energy density of the two-fluid system,
to which we will refer as the ``equation of state'' (EoS) in the
following. Using the normalization conditions of the 4-velocities 
\begin{equation}
\label{norm}
g_{\mu\nu}{u}_{\n}^{\ \mu}{u}_{\n}^{\ \nu} = -1 \ \ \text{and} \
\ g_{\mu\nu}{u}_{\p}^{\ \mu}{u}_{\p}^{\ \nu} = -1, 
\end{equation}
the components of the 4-currents read
\begin{equation}
\label{4courant}
n_{\n}^{\ \mu} = n_{\n} u_{\n}^{\ \mu} \ \ \text{and} \ \
n_{\p}^{\ \mu} = n_{\p} u_{\p}^{\ \mu}, 
\end{equation}
from which we interpret the quantity $n_{X}$ as the particle density
of the fluid $X$, as measured in its proper rest frame.
 
From variations of the Lagrangian density (keeping the metric fixed),
one defines the conjugate momenta $p_{\ \mu}^{\n}$ and
$p_{\ \mu}^{\p}$ as follows
 \begin{equation}
\label{impulsions}
\df \Lambda = p_{\ \mu}^{\n} \df n_{\n}^{\ \mu} + p_{\ \mu}^{\p} \df
n_{\p}^{\ \mu}. 
\end{equation}

Using (\ref{covariance}), these momenta are given in terms of the
4-currents by
\begin{equation}
\label{matrice}
\begin{pmatrix}
   p_{\ \mu}^{\n}   \\
   p_{\ \mu}^{\p}  
\end{pmatrix}
= 
\begin{pmatrix}
   \K^{\n\!\n} & \K^{\n\!\p}   \\
   \K^{\p\!\n} & \K^{\p\!\p} 
\end{pmatrix}
\begin{pmatrix}
   n_{\ \mu}^{\n}   \\
   n_{\ \mu}^{\p}  
\end{pmatrix}
\end{equation}
where $\K^{X\! Y}$ is the \textit{entrainment matrix}
\citep{andreev1976three}, whose components are defined from the EoS by
\begin{equation}
\label{defK1}
\K^{\n\!\n} = 2\left(\frac{\partial \E}{\partial
    n_{\n}^2}\right)_{n_{\p},x},\  \ \  \K^{\p\!\p} =
2\left(\frac{\partial \E}{\partial n_{\p}^2}\right)_{n_{\n},x},  
\end{equation}
\begin{equation}
\label{defK2}
  \K^{\n\!\p} = \K^{\p\!\n} = \left(\frac{\partial \E}{\partial x^2}\right)_{n_{\n},n_{\p}}.
\end{equation}

Because of the presence of the non-zero off diagonal term
$\K^{\n\!\p}$, the conjugate momentum of a fluid is not simply
proportional to its 4-velocity, but also depends on the 4-velocity of
the second fluid. This corresponds to the so-called
\textit{entrainment} effect.

To describe the difference in the fluid velocities, one introduces the
relative Lorentz factor $\Gamma_{\Delta}$  
\begin{equation}
\label{lorentzrel}
\Gamma_{\Delta} = - g_{\mu\nu} u_{\n}^{\ \mu} u_{\p}^{\ \nu} = \frac{x^2}{n_{\n}n_{\p}},
\end{equation}
to which we associate the relative speed $\Delta$ via 
\begin{equation}
\label{defDelta}
\Gamma_{\Delta} = \frac{1}{\sqrt{1-\Delta^2}}.
\end{equation}
$\Delta^2$ stands for the square of the physical speed of the protons
in the frame of neutrons (\ref{relspeed}), or the inverse. The
EoS~(\ref{covariance}) can be seen as a function of both densities and
the relative speed: $\E(n_{\n},n_{\p},\Delta^2)$. The first law of
thermodynamics then reads as
\begin{equation}
\label{thermo1bis}
\df\E = \mu^{\n}\df n_{\n} +\mu^{\p}\df n_{\p} + \alpha \df\Delta^2,
\end{equation}
where $\mu^{\n}$ and $\mu^{\p}$ denote neutron and proton chemical
potentials and $\alpha$ is the entrainment. The
$\K^{X\!  Y}$ elements are expressed as functions of these three
conjugate variables by
\begin{equation}
\label{K1}
\K^{\n\!\n}  = \frac{\mu^{\n}}{n_{\n}} -
\frac{2\alpha}{n_{\n}^2\Gamma_{\Delta}^2}, \ \ \K^{\p\!\p}  =
\frac{\mu^{\p}}{n_{\p}} - \frac{2\alpha}{n_{\p}^2\Gamma_{\Delta}^2},  
\end{equation}
\begin{equation}
\label{K2}
\K^{\n\!\p}  = \frac{2\alpha}{n_{\n}n_{\p}\Gamma_{\Delta}^3}. 
\end{equation}

The energy-momentum tensor $T_{\mu\nu}$ governing a mixture of two perfect
fluids is given by \citep{langlois1998differential}
\begin{equation}
\label{energieparfait2}
 T_{\mu\nu} = n_{\n\mu}p^{\n}_{\ \nu} + n_{\p\mu}p^{\p}_{\ \nu}+ \Psi g_{\mu\nu},
\end{equation}
where $\Psi$ is the generalized pressure of the system, linked to the
EoS through the Gibbs-Duhem relation 
\begin{equation}
\label{thermo2} 
\Psi(\mu^{\n},\mu^{\p}, \Delta^2)  = - \E +\, n_{\n}\mu^{\n}+ n_{\p}\mu^{\p},
\end{equation}
from which we get
\begin{equation}
\label{derthermo1}
n_{\n}  =\left( \frac{\partial \Psi}{\partial \mu^{\n}
  }\right)_{\mu^{\p},\Delta^2}, \ \ n_{\p}  = \left( \frac{\partial
    \Psi}{\partial \mu^{\p} } \right)_{ \mu^{\n},\Delta^2},
\end{equation}
\begin{equation}
\label{derthermo2}
\alpha = -\left( \frac{\partial \Psi}{\partial\Delta^2 } \right)_{ \mu^{\n},\mu^{\p}}.
\end{equation}

\subsection{Structure equations}
\label{section_structure}

In our study, we take the point of view of the 3+1 formalism
\citep{gourgoulhon20123+}, in which the spacetime $\M$ is foliated by
a family $\left( \Sigma_t \right)_{t \in \mathbb{R}}$ of spacelike
hypersurfaces. Let $\eul^\mu$ be the unit (future-oriented) vector
normal to $\Sigma_t $
\begin{equation}
\label{defn}
\eul^{\mu} = -N \nabla^{\mu} t  = \left(\frac{1}{N}, 0, 0,
  \frac{\omega}{N} \right). 
\end{equation}
As $\eul^\mu$ is a unit timelike vector, it can be seen as the
4-velocity of a given observer $\mathcal{O}_\eul$, called
\textit{Eulerian} or locally non-rotating observer.

In our choice of gauge~(\ref{metrique}), Einstein Equations form a set
of four coupled elliptic partial differential equations for the metric
potentials~\cite{bonazzola1993axisymmetric}. Matter
source terms involved in these equations are the energy density $E$,
the momentum density $\pi_\mu$ and the shear tensor $S_{\mu\nu}$
measured by $\mathcal{O}_\eul$. These quantities, which naturally
appear in the 3+1 decomposition of the energy-momentum tensor, are
defined by
\begin{equation}
\label{eulerien}
\left\{
  \begin{array}{rcl}
	E &=&  T_{\mu\nu}\eul^{\mu}\eul^{\nu}\\
	\pi_{\mu} &=& - T_{\rho\sigma}\eul^{\rho} {\gamma^{\sigma}}_{\mu} \\
	S_{\mu\nu}  &=& T_{\rho\sigma}{\gamma^{\rho}}_{\mu} {\gamma^{\sigma}}_{\nu}
  \end{array}
\right.
\end{equation}
where $\gamma_{\mu\nu}$ is the metric induced by $g_{\mu\nu}$ on the spacelike
hypersurface $\Sigma_t$. The matter source terms (\ref{eulerien}) are
functions of the entrainment matrix coefficients
(\ref{matrice}), the pressure $\Psi$, both densities and the
physical velocities measured by $\mathcal{O}_\eul$ (\ref{Un}).

The spacetime being circular (see Sec.~\ref{assumptions}), $u_{\n}^{\
  \mu}$ and $u_{\p}^{\ \mu}$ belong to the vector plane generated by
the two Killing vectors $\xi^\mu$ and $\chi^\mu$
\citep{bonazzola1993axisymmetric}. The angular velocities of the
fluids as seen by a static observer located at spatial infinity are
defined as follow
\begin{equation}
\Omega_{\n} = \frac{u_{\n}^{\ \varphi}}{u_{\n}^{\ t}} \ \ \text{and} \ \
\Omega_{\p} = \frac{u_{\p}^{\ \varphi}}{u_{\p}^{\ t}}.
\end{equation}
From these relations one defines $\Gamma_{\n}$ and $\Gamma_{\p}$, the Lorentz
factors of both fluids with respect to $\mathcal{O}_\eul$:
\begin{equation}
\label{defGamma}
\Gamma_{\n} = - \eul_\mu u_{\n}^{\ \mu} = N u_{\n}^{\ t} \ \
\text{and} \ \ \Gamma_{\p} = - \eul_\mu u_{\p}^{\ \mu}= N
u_{\p}^{\ t}.  
\end{equation}

We define $U_{\n}$ and $U_{\p}$ as the norms of the physical
3-velocities of the fluids measured by the Eulerian observer
$\mathcal{O}_\eul$, \textit{i.e.}
\begin{equation}
\label{Un}
U_{\n} = \frac{B}{N}(\Omega_{\n}-\omega) r\sin\theta \ \ \text{and} \
\ U_{\p} = \frac{B}{N}(\Omega_{\p}-\omega) r\sin\theta. 
\end{equation}
The normalization conditions on $\eul^\mu$, $u_{\n}^{\ \mu}$
and $u_{\p}^{\ \mu}$ lead to the standard expressions: 
\begin{equation}
\label{GammaExpr}
\Gamma_{\n} = \left(1-U_{\n}^2\right)^{-1/2}  \ \ \text{and} \ \
\Gamma_{\p} = \left(1-U_{\p}^2\right)^{-1/2} . 
\end{equation}

Moreover, the relative speed $\Delta$ (\ref{defDelta}) can be
expressed in terms of $U_{\n}$ and $U_{\p}$, by  
\begin{equation}
\label{relspeed}
\Delta^2=\frac{\left(U_{\n} - U_{\p} \right)^2}{\left(1-U_{\n}U_{\p}\right)^2}. 
\end{equation}

The equations governing the fluid equilibrium are derived from the
conservation of both particle 4-currents
\begin{equation}
  \label{partcons}
  \nabla_ {\mu} n_{\n}^{\ \mu} = 0 \ \ \text{and} \ \ \nabla_{\mu}
  n_{\p}^{\ \mu} = 0,
\end{equation}
which are trivially satisfied given the symmetries of the spacetime,
and from $\nabla^{\mu} T_{\mu\nu}=0$, the energy-momentum conservation
law. In the case of rigid rotation that we are considering here (see
Sec.~\ref{assumptions}), it leads to the two following first integrals
of motion
\begin{equation}
  \label{intprem}
  \frac{\mu^{\n}}{\Gamma_{\n}}N = \tilde{C}_{\n} \ \ \text{and} \ \
  \frac{\mu^{\p}}{\Gamma_{\p}}N = \tilde{C}_{\p}, 
\end{equation}
where $\tilde{C}_{\n}$ and $\tilde{C}_{\p}$ denote constants over the
whole star. Introducing the log-enthalpies
\begin{equation}
  \label{logEnthal}
  H^{\n}= \ln\left(\frac{\mu^{\n}}{m_{\n}}\right) \ \ \text{and} \ \
  H^{\p}= \ln\left(\frac{\mu^{\p}}{m_{\p}}\right), 
\end{equation}
with $m_{\n}=939.6$ MeV and $m_{\p}=938.3 + 0.5 = 938.8$ MeV the
masses of particles composing the fluids, one can rewrite
(\ref{intprem}) as 
\begin{equation}
  \label{intprem2}
  H^{\n}+\ln N-\ln\Gamma_{\n} = C_{\n} \ \ \text{and} \ \ H^{\p}+\ln
  N-\ln\Gamma_{\p} = C_{\p}, 
\end{equation}
$C_{\n}$ and $C_{\p}$ being constant over the star.

In Section IV, we will only present configurations verifying chemical
equilibrium at the center of the star, \textit{i.e.}
\begin{equation}
  \label{equality1}
  \mu^{\p}_c= \mu^{\n}_c,
\end{equation}
or equivalently, 
\begin{equation}
  \label{equality2}
  H^{\n}_c= H^{\p}_c+ \ln \left(\frac{m_{\p}}{m_{\n}}\right).
\end{equation}
Putting (\ref{equality1}) in (\ref{intprem}), one gets $\tilde{C}_{\n}
= \tilde{C}_{\p}$. Inside the star, the chemical potentials are thus
linked through
\begin{equation}
  \label{eqrot}
  \frac{\mu^{\n}}{\Gamma_{\n}} = \frac{\mu^{\p}}{\Gamma_{\p}}.
\end{equation}

As shown by Andersson and Comer~\cite{andersson2001slowly}, global
$\beta$-equilibrium is only possible if the two fluids are
corotating. In this case, imposing chemical equilibrium at the center
of the star is enough for the chemical equilibrium to be verified in
the whole star, as can be seen from (\ref{eqrot}). In the opposite
case, where $\Omega_{\n} \neq \Omega_{\p}$, some conversion reactions
between neutrons and protons should be included in our model, which
would dissipate some energy until the star reaches $\beta$-equilibrium
with $\Delta^2 = 0$ \citep{prix2002slowly}. However, as we are dealing
with stationary configurations, this transfusive process is neglected
(see Sec.~\ref{assumptions} and (\ref{partcons})).  This assumption
makes sense because of the slowness of the electroweak reactions
responsible for the chemical equilibrium \citep{yakovlev2001neutrino},
added to the fact that the two fluids are likely to be always very
close to corotation\footnote{Assuming the total angular momentum to be
  constant during a glitch, the maximum lag between the fluids, which
  corresponds to the lag when the glitch is triggered, is roughly
  given by $\Delta \Omega_{\text{max}} \simeq I / I_{\n} \times \Delta
  \Omega_{\p} / \Omega_{\p} \times \Omega \simeq \Delta \Omega_{\p} /
  \Omega_{\p}\times \Omega$, where $\Delta \Omega_{\p} /
  \Omega_{\p}\sim 10^{-11} - 10^{-5}$ is the glitch amplitude and
  $\Omega$ is the pulsar angular velocity
  (e.g. \citep{sidery2010dynamics}).}. Examples of configurations with
$\mu^{\p}_c \neq \mu^{\n}_c$ are shown in Prix \textit{et
  al.}~\citep{prix2005relativistic}.

\subsection{Global quantities}\label{global_quantities}

We give here some definitions which we use in Section~\ref{eqconf};
more details are given in Prix \textit{et
  al.}~\cite{prix2005relativistic}. The \textit{gravitational mass}
($M_G$) is the mass felt by a test-particle orbiting around the
star. It is defined as the (negative) coefficient of the term $1/r$ in
an asymptotic expansion of the $\log N$ gravitational
potential. Following Bonazzola \textit{et
  al.}~\cite{bonazzola1993axisymmetric}, it can be expressed as
\begin{equation}
  M_G = \int_{\Sigma_t} \left[ N \left(E + S_i^{\ i} \right) + 2B^2
    r^2 \sin^2\theta\, \omega \pi^\varphi  \right] \df^{\, 3}\!
  \Sigma  \label{e:def_mass_grav},
\end{equation}
where $\df^{\, 3}\! \Sigma = A^2Br^2\sin \theta \df r \df \theta \df
\varphi$ is the element volume on the hypersurface $\Sigma_t$. The
\textit{baryon mass} ($M^B$) is nothing but the counting of the total
number of baryons in the star. In our case, it splits into two parts:
neutron baryon mass ($M^B_{\n}$) and proton baryon mass ($M^B_{\p}$).

Relying on the axisymmetry of the spacetime, associated with the
Killing vector $\chi^\mu$ (cf. Sec.~\ref{assumptions}), the total
angular momentum of the star is given by the gauge-invariant Komar
formula \citep{komar1959covariant}
\begin{equation}
\label{defKomarVol}
J_{\text{K}} = - \int_{\Sigma_{t}}
\eul^\mu\,T_{\mu\nu}\,\chi^\nu \sqrt {\gamma} \mathrm{d}^3 x, 
 \end{equation}
 where $\gamma$ is the determinant of the 3-metric $\gamma_{ij}$ defined as the restriction of the metric $\gamma_{\mu\nu}$ to the hypersurface $\Sigma_t$ (see Sec.~\ref{section_structure}), such that $\gamma_{ij} = g_{ij}$ (cf. Eq.~(\ref{metrique})). From (\ref{eulerien}), we deduce that
 $\eul^\mu\,T_{\mu\nu}\,\chi^\nu = - \pi_{\varphi}$, so that
(\ref{defKomarVol}) is simply given by \citep{gourgoulhon20123+}
\begin{equation}
\label{defKomarVolCOOL}
J_{\text{K}} = \int_{\Sigma_{t}} \pi_{\varphi}\  \df^{\, 3}\! \Sigma. 
 \end{equation}
 For a two-fluid system (\ref{energieparfait2}), we can write:
 \begin{equation}
   \label{dec}
   \pi_{\varphi} = \Gamma_{\n}n_{\n}p^{\n}_{\varphi} + \Gamma_{\p}n_{\p}p^{\p}_{\varphi},
 \end{equation}
 see Eqs.~(\ref{impulsions}) and (\ref{defGamma}). Note that there is
 no term involving the pressure $\Psi$. This canonical decomposition
 leads us to define the angular momentum density of each fluid as in
 \citep{langlois1998differential}
 \begin{equation}
   j^{\n}_{\varphi} \equiv \Gamma_{\n}n_{\n}p^{\n}_{\varphi} \ \
   \text{and} \ \ j^{\p}_{\varphi} \equiv
   \Gamma_{\p}n_{\p}p^{\p}_{\varphi}. 
 \end{equation}
One can thus interpret $p^{\n}_{\varphi}$ (resp. $p^{\p}_{\varphi}$)
as the angular momentum per neutron (resp. proton) and
$\Gamma_{\n}n_{\n}$ (resp. $\Gamma_{\p}n_{\p}$) as the density of
neutrons (resp. protons) measured by $\mathcal{O}_\eul$, $n_{\n}$
(resp. $n_{\p}$) being the density of neutrons (resp. protons) in the
frame of this fluid. These angular momentum densities are expressible
as functions of the two physical velocities measured by
$\mathcal{O}_\eul$ (\ref{Un})
\begin{equation}
\label{def1}
\left\{
  \begin{array}{rcl}
	j^{\n}_{\varphi} = Br\sin\theta
        (\Gamma_{\n}^2n_{\n}^2\K^{\n\!\n}U_{\n} +
        \Gamma_{\n}n_{\n}\Gamma_{\p}n_{\p}\K^{\n\!\p}U_{\p}),  \\[3 pt]
	j^{\p}_{\varphi} = Br\sin\theta
        (\Gamma_{\p}^2n_{\p}^2\K^{\p\!\p}U_{\p} +
        \Gamma_{\n}n_{\n}\Gamma_{\p}n_{\p}\K^{\n\!\p}U_{\n}). 
  \end{array}
\right.
\end{equation}

Using Eqs. (\ref{defKomarVolCOOL}) and (\ref{dec}), we deduce that the
angular momentum of each fluid is given by
\begin{equation}
  \label{def2}
  J_{\n} = \int_{\Sigma_{t}} j^{\n}_{\varphi}\ \df^{\, 3}\! \Sigma \ \
  \text{and} \ \ J_{\p} = \int_{\Sigma_{t}} j^{\p}_{\varphi}\ \df^{\,
    3}\! \Sigma .
\end{equation}
The Newtonian limit of the angular momenta is studied in
appendix~\ref{Jnewt} and compared to results from Sidery \textit{et
  al.}~\citep{sidery2010dynamics}.

Assuming rigid rotation, from the fluid angular momenta it is possible
to define corresponding moments of inertia. The total moment of
inertia of the star is
\begin{equation}
I = \frac{J}{\Omega_{\p}},
\end{equation}
$\Omega_{\p}$ corresponding to the rotation rate of the pulsar. The
moment of inertia of fluid $X$ can be defined through the equation
\begin{equation}
I_X = \frac{J_X}{\Omega_{X}},
\end{equation}
which makes sense if the two fluids are corotating\footnote{In the
  general relativistic framework, there is no natural decomposition of
  $J_X$ in the form of Eq. (\ref{newto}). By assuming $\Delta^2=0$, we
  ensure that $I_{\n} + I_{\p} = I$.}.

\subsection{Numerical procedure}
\label{procedure}

The numerical resolution of the stationary axisymmetric configurations
described in the previous sections was implemented in the
\textsc{lorene} library by Prix \textit{et
  al.}~\cite{prix2005relativistic}. It is based on an iterative
scheme, called self-consistent-field method, which consists in making
an initial guess on the quantities to be determined, starting from a
flat spacetime with both fluids at rest and parabolic profiles for
$H^{\n} \left(r, \theta \right)$ and $H^{\p} \left(r, \theta \right)$,
and progressively improving these estimates at each step of the
resolution procedure, until a convergence criterion is satisfied. For
a given EoS, the free parameters are the central values $H^{\n}_c$ and
$H^{\p}_c$ of the log-enthalpies and the (constant) angular velocities
$\Omega_{\n}$ and $\Omega_{\p}$; thus every set of such parameters
gives a model of rotating two-fluid neutron star.

Numerical techniques are based on multi-domain spectral methods
\citep{grandclement2009spectral}, which make it possible to reach a
high accuracy with a small number of coefficients. In the cold
single-fluid case \citep{bonazzola1993axisymmetric}, the surface of
the star is defined as the location where the pressure, or
equivalently the log-enthalpy, of the fluid is vanishing. For a
two-fluid system, it is not possible to define the surface of the
inner fluid with a vanishing log-enthalpy any more, because of the
coupling between both fluids (see appendix~\ref{a:num}). Instead, both
surfaces are taken to be the location where the corresponding density
vanishes, \textit{i.e.} $n_X = 0$
\citep{prix2005relativistic}. Consequently, our models assume that
both fluids are present at the center of the star, then one of them
vanishes (its density reaching zero), and there is a region with only
one fluid left, until this one disappears, too, defining the surface
of the star. In realistic configurations, for which $\Omega_{\n}
\simeq \Omega_{\p}$ and $\mu^{\n} \simeq \mu^{\p}$
(cf. (\ref{eqrot})), the surfaces of the two fluids are very close to
each other, leading the region between both surfaces, with one fluid,
to be poorly represented by the grid covering the star. To cope with
this problem, we take one additional domain with many grid points to
represent the thin shell where the transition from two fluids to one
fluid and vacuum occurs. This solution happened to lead to a
significant improvement of the determination of the surfaces and on
the accuracy of the results
\citep{prix2005relativistic}. Consequently, four different domains are
used to cover the entire space in general: the innermost domain covers
the core of the star, the second one represents the outer part of the
star, a third one is used outside the star, expanding up to a few
stellar radii, and a last one describes the remaining part, up to
infinity with the help of a change in coordinates of the type $r=
\frac{1}{A(1-\zeta)}$ with $\zeta \in[-1;1]$.

\section[]{Equations of state}
\label{EOS}

\subsection[]{Presentation}

Although non-relativistic models are sufficient to describe the cores
of low-mass neutron stars \citep{chamel2008two}, a (special)
relativistic formulation, besides being self-consistent, is necessary
to deal with massive neutron stars. On the scales relevant for the
thermodynamic averaging leading to the equation of state, the metric
can be considered as (locally) flat
\cite{glendenning2000compact}. Therefore, within this section we will
work with a Minkowski metric, $\eta^{\mu\nu}$. For the
$\gamma$-matrices, we will use the anticommutation relation
$\{\gamma^\mu,\gamma^\nu\} = 2 \eta^{\mu\nu}$. The effect of
superfluidity/superconductivity on the EoS itself has been neglected
since pairing and superfluidity/superconductivity is a Fermi surface
effect with only a marginal influence on the EoS.

We will employ here two equations of state based on a phenomenological
relativistic mean field (RMF) model. This type of models can be
considered as realistic in the sense that they aim to describe as well
as possible known properties of finite nuclei and nuclear matter. The
basic idea is that the interaction between baryons is mediated by
meson fields inspired by the meson exchange models of the
nucleon-nucleon interaction. Within RMF models, these are, however,
not real mesons, but introduced on a phenomenological basis with their
quantum numbers in different interaction channels. The coupling
constants are adjusted to a chosen set of nuclear observables. Earlier
models introduce non-linear self-couplings of the meson fields in
order to reproduce correctly nuclear matter saturation properties,
whereas more recently density-dependent couplings between baryons and
the meson fields have been widely used. The literature on those models
is large and many different parametrizations exist (see
e.g.~\cite{dutra2014relativistic}).

In the present paper, we will use models with density dependent
couplings. The microscopic Lagrangian density of that type of models
can be written in the following form
\begin{eqnarray} 
  {\mathcal L} &=& \sum_{X =(n,p)} -\bar \psi_X \Big( 
    \gamma_\mu \partial^\mu + m_X - g_{\sigma} \sigma \Big. \nonumber   \\ 
  &&\left. - g_{\delta } \vec{\delta}\! \cdot\! \vec{I}_X - i g_{\omega}
    \gamma_\mu \omega^\mu - i  g_{\rho} \gamma_\mu \vec{\rho}^{\, \mu}\! \cdot\!
    \vec{I}_X \right) \psi_X \nonumber \\ 
    && - \frac{1}{2}  \left(\partial_\mu \sigma \partial^\mu \sigma + m_\sigma^2
  \sigma^2\right)\nonumber \\ 
  && - \frac{1}{2} \left(\partial_\mu   \vec{\delta} \partial^\mu \vec{\delta} + m_{\delta}^2   {\vec{\delta}}^{\, 2}\right) \nonumber \\ 
  && - \frac{1}{4} W^\dagger_{\mu\nu} W^{\mu\nu} - \frac{1}{2} m^2_\omega \omega_\mu
  \omega^\mu  \nonumber \\ 
  && - \frac{1}{4} \vec{R}^\dagger_{\mu\nu}\! \cdot\!   \vec{R}^{\mu\nu} - \frac{1}{2} m^2_\rho \vec{\rho}_{\, \mu} \!\cdot\!   \vec{\rho}^{\, \mu} ~.
\end{eqnarray}

Here, $\psi_X$ denotes the field of baryon $X$\footnote{Here
  ``$(n,p)$'' refers to particles (neutrons, protons), not
  fluids. Electrons shall be considered later in this Section.} with
rest mass $m_X$. The corresponding isospin operator is
$\vec{I}_X$. $W^{\mu\nu}$ and $\vec{R}^{\mu\nu}$ are the vector meson
field tensors of the form
\begin{eqnarray}
V^{\mu\nu} = \partial^\mu  V^\nu - \partial^\nu V^\mu~,
\end{eqnarray}
associated with $\omega^\mu$ and $\vec{\rho}^{\, \mu}$
respectively. $\sigma$ is a scalar-isoscalar meson field and
$\vec{\delta}$ induces a scalar-isovector coupling to differentiate
proton and neutron effective masses (\ref{eq:masses}). For $M$
spanning over all meson types $(\sigma, \rho, \delta, \omega)$, the
quantity $g_M$ stands for the coupling between nucleons and meson $M$,
whose mass is $m_M$.

We will show results within two density-dependent models,
DDH~\cite{typel1999relativistic} and
DDH$\delta$~\cite{gaitanos2004lorentz, avancini2004instabilities,
  avancini2009nuclear}. The $\delta$-field is absent in DDH. The
couplings are density dependent,
\begin{equation}
g_M(n_B) = g_M(n_0) h_M(x)~,\quad x = n_B/n_0~.
\end{equation}
$n_0$ thereby denotes a normalization constant, in most cases it is
chosen as the saturation density of symmetric nuclear matter. The
baryon number density $n_B$ is a scalar quantity defined as $n_B =
\sqrt{-n_B^{\ \mu} n_{B \mu}}$, where $n_B^{\ \mu} = n_{\p}^{\ \mu} +
n_{\n}^{\ \mu}$ is the total baryon current.

Within both parametrizations employed in this paper, the following
forms \citep{avancini2009nuclear} are assumed for the isoscalar
couplings $(M=\sigma, \omega)$
\begin{equation}
h_M(x) = a_M \frac{1 + b_M ( x + d_M)^2}{1 + c_M (x + d_M)^2}
\end{equation}
and 
\begin{equation}
h_M(x) = a_M\,\exp[-b_M (x-1)] - c_M (x-d_M)
\end{equation}
for the isovector ones $(M=\rho, \delta)$.  

\subsubsection{Single-fluid case}
In mean field approximation, the meson fields are replaced by their
respective mean-field expectation values \citep{haensel2007neutron,
  glendenning2000compact}. Assuming that all particles move at the
same speed, \textit{i.e.} for the single fluid case, in uniform matter
the following (Euler-Lagrange) relations emerge
\begin{subequations}
\begin{eqnarray}
m_\sigma^2 \bar\sigma &=& g_{\sigma} (n_{\p}^s + n_{\n}^s)
\\
m_\delta^2 \bar\delta &=& g_{\delta} (n_{\p}^s - n_{\n}^s)\\
m_\omega^2 \bar\omega &=& g_{\omega} (n_{\p} + n_{\n})\\
m_\rho^2 \bar\rho &=& g_{\rho} (n_{\p} - n_{\n})~,
\end{eqnarray}
\end{subequations}
where $\bar\sigma = \langle \sigma \rangle$,
$\bar\delta=\langle\delta_3\rangle$, $\bar\rho=\langle\rho_3^0\rangle$
and $\bar\omega=\langle\omega^0\rangle$. Note that only the isospin
3-components of the isovector meson fields contribute and, since the
fluid rest frame is chosen for convenience, only the 0-components of
the vector meson fields are non-vanishing
\citep{glendenning2000compact}. The scalar density of baryon $X$ is
given by
\begin{equation}
n^s_X = \langle \bar \psi_X \psi_X \rangle = 2 \int
f(e_X(k_\nu))\frac{d^3 k}{(2 \pi)^3} \frac{m^*_X} {e_X(k_\nu)}~,
\end{equation}
and the number density by 
\begin{eqnarray}
n_X &=& n_X^0 = i \,\langle \bar \psi_X\gamma^0 \psi_X \rangle
\nonumber \\ &=& 2 \int f(e_X(k_\nu)) \frac{d^3k}{(2 \pi)^3} =
\frac{\left(k_{F,X}\right)^3}{3 \pi^2} ~, 
\label{eq:density}
\end{eqnarray}
where $k_{F,X}$ is the Fermi momentum of fluid $X$. $f$ represents
here the fermionic distribution function with single-particle energies
$e_X$. Note that the distribution function is a scalar quantity. At
zero temperature, this is a Heavyside step function equal to 1 for
occupied states (corresponding to $k\leq k_{F,X}$) and 0 for
non-occupied ones. The argument can be written in a covariant way as
$\mu^X + k_\nu u^\nu$, where $k_\nu$ represents the (on-shell)
momentum of a single particle state and $u^\nu$ the four-velocity of
the actual reference frame. For the single-fluid case, where the fluid
rest frame can be chosen as reference frame, this reduces to the well
known form $ f(e_X) = \theta(\mu^X_* - e_X)$ with
\begin{equation}
e_X(k_\nu) = \sqrt{k^i k_i + (m^*_X)^2}.
\end{equation}  The Dirac effective masses $m^*_X$ depend on the scalar
mean fields as
\begin{equation}
m^*_X = m_X - g_{\sigma} \bar\sigma -
g_{\delta} t_{3 X} \bar \delta~,
\label{eq:masses}
\end{equation}
where $t_{3 X}$ indicates the third component of isospin, with the
convention $t_{3 \p} = 1 $ and $t_{3 \n} = -1 $.  The effective
chemical potentials $\mu^X_*$, also called Landau effective masses
\citep{gusakov2009relativistic, urban2015collective}, are defined as
\begin{equation}
\mu^X_*~=\sqrt{\left(m_X^*\right)^2~+~\left(k_{F,X}\right)^2}.
\end{equation}
In the single fluid case, these quantities are related to the chemical
potentials via \citep{avancini2009nuclear}
\begin{subequations}
\begin{eqnarray}
\mu^{\n} &=& \mu^{\n}_* + a_+ n_{\n} + a_- n_{\p} + \Sigma^R~ \label{eq:mu1}\\
\mu^{\p} &=& \mu^{\p}_* + a_+ n_{\p} + a_- n_{\n} + \Sigma^R~ \label{eq:mu2}
\end{eqnarray} 
\end{subequations}
with $a_\pm = g_\omega^2/m_\omega^2 \pm g_\rho^2/m_\rho^2$. 
The rearrangement term 
\begin{eqnarray}
\Sigma^R &=& \frac{\partial g_{\omega}}{\partial n_B} \frac{g_\omega}{m_\omega^2} n_B^2 + \frac{\partial g_{\rho}}{\partial n_B} \frac{g_\rho}{m_\rho^2} n_I^2 \nonumber \\ &&  -\frac{\partial g_{\sigma}}{\partial n_B}
\bar\sigma (n_{\p}^s + n_{\n}^s) - \frac{\partial g_{\delta}}{\partial n_B}
\bar\delta (n_{\p}^s - n_{\n}^s) ~.
\label{eq:sigmar}
\end{eqnarray}
is present in density-dependent models to ensure thermodynamic consistency. 
We have used here the definition of the baryon number density $n_B = n_{\p} + n_{\n}$
and have introduced the isospin density $n_I = \sqrt{-n_I^{\ \mu} n_{I \mu}}$, where $n_I^{\ \mu} = n_{\p}^{\ \mu} - n_{\n}^{\ \mu}$.

The wealth of nuclear data allows to constrain reasonably the
parameter values of the interaction between nucleons. The
corresponding parameter values of both models can be found in the
above references \citep{typel1999relativistic, avancini2009nuclear}
and the resulting nuclear matter properties are listed in
Table~\ref{tab:nuclear}. The two models differ only in the isovector
channels, thus the properties of symmetric nuclear matter are
similar. For the EoS of compact stars, the isospin dependence of the
EoS is extremely important since very asymmetric matter close to pure
neutron matter is encountered.  The two quantities containing
information about the isospin dependence of the EoS are the symmetry
energy $J$ and its slope $L$ at saturation density. Another
interesting quantity in this respect is the EoS of pure neutron matter
at low densities, where recent progress in microscopic calculations
has allowed to obtain valuable constraints. In
\citep{kruger2013neutron}, a range
\begin{equation}
14.1 \lesssim E/A(n_0)\lesssim 21.0 \mathrm{\ MeV}
\label{Kruger}
\end{equation}
has been derived for the energy per baryon of pure nuclear matter
(neutron mass subtracted) from microscopic calculations within chiral
nuclear forces. The corresponding value within the two models used
here is given in Table~\ref{tab:nuclear}, too. 

Saturation properties of symmetric nuclear matter are in reasonable
agreement with nuclear data~\citep{danielewicz2009symmetry,
  piekarewicz2010we}. As can be seen within the DDH$\delta$-model, the
symmetry energy and its slope lie at the lower end of reasonable
values (cf. \citep{tsang2012constraints, lattimer2013constraining,
  lattimer2014constraints} for a compilation and discussion of
constraints obtained from nuclear experiments) and the energy per
baryon of pure nuclear matter is probably too low, too. Within DDH the
values are much larger, indicating a much stiffer EoS in strongly
asymmetric matter. The choice of these two models therefore allows to
explore different interactions in the equilibrium configurations
presented here.

\begin{table*}[!t]
  \caption{ \label{tab:nuclear}
    Nuclear matter properties at saturation density of the
    two models considered in this study. $n_0$ thereby denotes the
    saturation density, $B_{\text{sat}}$ the binding energy, $K$ the
    incompressibility, $J$ the symmetry energy, $L$ the slope of the
    symmetry energy and $E/A(n_0)$ is the energy per baryon of pure
    neutron matter with the neutron mass subtracted, see
    e.g.~\citep{typel2015compose} for a 
    definition of the different quantities. The maximum
    gravitational masses of neutron stars assuming corotation and
    $\beta$-equilibrium, see Sec.~\ref{eqconf}, are given, too.}
\begin{ruledtabular}
\begin{tabular}{c|cccccccc}
& $n_0 $ & $B_{\text{sat}}$ & $K$& $J$    &$L$  &$E/A\ (n_0)$& $M_G^{\mathit{max}}\ (0\  \mathrm{Hz})$ & $M_G^{\mathit{max}}\ (716\ \mathrm{Hz})$  \\[2 pt]  
& $[\ \mathrm{fm}^{-3}\ ]$   & [ MeV ] & [ MeV ] & [ MeV ] & [ MeV ] & [ MeV ]&
      [ $M_\odot$ ] & [ $M_\odot$ ]\\[2 pt]    \hline \\[-5 pt]
      \ \ \ \textbf{DDH} \ \ \  & 0.153 & 16.3 &  240 & 33.4   & 55 &18.4 &2.08 &2.12 \\[2 pt]
      \textbf{DDH}\bm{$\delta$} & 0.153 & 16.3  & 240&25.1 &44 & 10.6 & 2.16 & 2.21\\
\end{tabular}
\end{ruledtabular}
\end{table*}

\subsubsection{Two-fluid case}
In a two-fluid system, no common rest frame for both fluids can be
defined and the system's equation of state becomes a function of the
relative speed $\Delta$ between both fluids. In non-relativistic
models, commonly the Fermi liquid theory is employed to calculate the
(Andreev-Bashkin) entrainment matrix, see
e.g.~\cite{chamel2006entrainment, chamel2008two}. For relativistic
two-fluid systems, two different approaches can be found in the
literature. On the one hand, Gusakov et
al.~\cite{gusakov2009relativistic, gusakov2014physics} have used a
relativistic generalization of Fermi liquid theory to calculate the
entrainment matrix of homogeneous matter containing, in addition to
electrons, nucleons or more generally the whole baryon octet. Results
from this approach within a density-dependent model can be found in
\cite{urban2015collective}. On the other hand,
\cite{comer2003relativistic} have presented a formalism to evaluate
the master function $\Lambda$ from the thermodynamic average (at
mesoscopic scales) of the energy-momentum tensor and applied it to a
simple RMF model containing only isoscalar interactions. The
entrainment matrix can then be evaluated from the derivatives,
following the definitions in Sec.~\ref{hydro}. The same formalism has
been applied to a more advanced and more realistic RMF model with
isovector interaction by Kheto and
Bandyopadhyay~\cite{kheto2014isospin}.

Here, we will follow the strategy of \cite{comer2003relativistic} and
show that the resulting entrainment matrix is in agreement with that
obtained from relativistic Fermi liquid theory in the limit of small
relative speed as it should be. Our aim is to calculate the master
function $\Lambda$ which is a scalar quantity, depending on the three
scalars, $n_{\n}, n_{\p},\Delta^2$. For convenience, we choose the
zero-velocity frame of the neutron fluid (see Sec~\ref{sub:entr}) in
which the proton fluid acquires a nonzero three-velocity,
$v^i$. Without loss of generality we can choose $v^i$ to be oriented
in $z$-direction in order to simplify the computations, \textit{i.e.}
$v^i = (0,0,v)$.

Following \citep{comer2003relativistic}, the master function reads as
\begin{equation}
\Lambda = -\langle \tau^{00}\rangle  - \langle \tau^{xx}\rangle + \langle \tau^{zz}\rangle~,
\end{equation}
where $\langle \tau^{\mu\nu} \rangle = T^{\mu\nu}$
(\ref{energieparfait2}) corresponds to the thermal expectation value
of the elements of the energy-momentum tensor. Neglecting gradients of
the mesonic mean fields, the microscopic energy-momentum tensor can be
written as
\begin{equation}
\tau^{\mu\nu} = \sum_X \frac{1}{2}( \bar{\psi}_X\gamma^\mu \partial^\nu \psi_X +
(\partial^\mu \bar{\psi}_X) \gamma^\nu \psi_X) + g^{\mu\nu} {\mathcal
  {L}}~. 
\end{equation}
The particle currents are given by $n_X^{\ \nu} = n_X u_X^{\ \nu} = i
\langle \bar{\psi}_X\gamma^\nu \psi_X\rangle$. Since we have chosen
the zero-velocity frame of the neutron fluid, only the proton current
has nonzero spatial components with
\begin{equation}
\label{eq:expected}
n_{\p}^{\ \nu} = \frac{n_{\p}}{\sqrt{1-v^2}}(1,0,0,v).
\end{equation}
Due to the nonzero proton velocity, the mean fields of the vector
mesons acquire nonzero spatial components, too, following the
relations:
\begin{subequations}
\begin{eqnarray}
m_\omega^2 \langle\omega^i\rangle &=& g_{\omega} \left(n_{\p}^{\
    i} + n_{\n}^{\ i}\right) = g_{\omega} \, n_B^{\ i}\\
m_\rho^2 \langle\rho^i\rangle &=& g_{\rho} \left(n_{\p}^{\
  i}-n_{\n}^{\ i}\right) = g_{\rho}\, n_I^{\ i}~,
\end{eqnarray}
\end{subequations}
where $\langle \rho^i \rangle \equiv \langle \rho^i_3 \rangle$. For
better readability we will suppress the brackets for the mean field
expectation values of the meson fields in the following equations. In
addition, since we have chosen the fluid velocity in $z$-direction,
only the $z$-components become nonzero.

Let us now check that indeed the resulting proton and neutron currents
have the assumed form, with $n_X$ given by the respective rest frame
expressions, $k_{F,X}^3/(3 \pi^2)$. The following derivations differ
slightly from that exposed in~\cite{comer2003relativistic,
  kheto2014isospin}. In \cite{comer2003relativistic,
  kheto2014isospin}, in order to account for the moving proton fluid,
the Fermi momentum of protons has been shifted by a momentum $K$,
whereas that of the neutrons has been kept the same with the argument
that the reference frame is the neutron zero spatial momentum
frame. However, following this strategy, the relativistic deformation
of the Fermi sphere, which shows up at second order in the velocities,
is not taken into account. In our opinion, this is the reason why the
final result for the entrainment matrix in
\cite{comer2003relativistic, kheto2014isospin} does not agree with the
Fermi liquid theory result~\cite{gusakov2009relativistic}.  Therefore,
we will use a different method~\cite{baym1976landau}, namely we will
use the Lorentz transformation properties of the different involved
quantities to calculate the master function in the neutron rest frame,
but where the proton fluid has nonzero spatial velocity. An advantage
of this method is that it allows to calculate the master function to
any order in the velocity and that the deformation of the Fermi sphere
is automatically included. Note, however, that we do not include
  any velocity-dependent modification of the superfluid energy gap and
  that thus our results can be applied only for relative velocities
  below the superfluid critical velocity, which should be of the order
  of $10^7$ cm.s$^{-1}$ in neutron stars~\cite{Gusakovvelocity}.

Let us start with the zero components, $n_X^0 =
i \,\langle\bar\psi_X\gamma^0\psi_X\rangle$. Due to the nonzero value of
the spatial components of the mesonic mean fields, the single particle
kinetic energies are modified and become 
\begin{eqnarray}
e_X(k_\nu) &=& \sqrt{
(k^z - g_{\omega} \omega^z - g_{\rho} t_{3 X} \rho^z)^2 +
(m^*_X)^2} \nonumber \\
&\equiv& \sqrt{ k'_ik'^i + (m^*_X)^2}.
\end{eqnarray}
For the neutrons, since we are in the zero-velocity frame, a
simple shift in the integration variable $k_i \to k'_i$ shows
that $n_{\n}^{\ 0} = n_{\n}$ as it should be. For the protons, since the proton
fluid has a nonzero velocity, all momenta are Lorentz boosted, \textit{i.e.}
\begin{equation}
n_{\p}^{\ 0} = 2 \int \frac{d^3 \tilde{k}}{(2 \pi)^3}
f\left(\tilde{e_{\p}}\left(\tilde{k}_\nu\right)\right),
\end{equation}
where the quantities in the moving frame have been denoted by a tilde.
Using the fact that the distribution function is a scalar with a
scalar argument, and that $k^\alpha$ transforms as a vector under
Lorentz transformations, we can express the integrand with quantities
in the zero-velocity frame of the protons (see
e.g.~\cite{baym1976landau})
\begin{equation}
n_{\p}^{\ 0} = 2 \int \frac{d^3 k}{(2 \pi)^3} J(k,\tilde{k})
\theta\left(\mu^{\p}_* - e_{\p}(k_\nu)\right) ~.
\end{equation}
$J(k,\tilde{k})$ denotes here the Jacobian for the change in
integration variable from $d^3\tilde{k} \to d^3k$, which is given by 
\begin{equation}
J(k,\tilde{k}) = \frac{1}{\sqrt{1-v^2}} \, \left(1 + v \,
  \frac{\partial e_{\p}(k_\nu)}{\partial k^z}\right)~. 
\end{equation}
Evaluating the integration leads to the desired result, $n_{\p}^0 =
n_{\p}/\sqrt{1-v^2} = n_{\p} u_{\p}^0$.

Similarly, the $z$-components of the currents can be evaluated, with
\begin{eqnarray}
  n_{\n}^z &=& \int d^3k\, \theta(\mu^{\n}_* - e_{\n}(k^\nu)) \frac{k^z
    - g_\omega \omega^z - g_\rho \rho^z}{e_{\n}(k^\nu)} \nonumber \\ 
    &=& 0 \\ 
  n_{\p}^z &=& \int d^3k\, \theta(\mu^{\p}_* - e_{\p}(k^\nu)) 
  J(k,\tilde{k}) \nonumber \\ 
  && \times \frac{  e_{\p}(k^\nu) v + k^z  - g_\omega \omega^z + g_\rho
    \rho^z}{e_{\p}(k^\nu) + v (k^z - g_\omega \omega^z + g_\rho
    \rho^z)}  \nonumber \\ 
  & =& n_{\p} \frac{v}{\sqrt{1-v^2}}~. 
\end{eqnarray}
This is indeed the expected result (\ref{eq:expected}). 

Let us now turn to the evaluation of the master function. After some
algebraic manipulations and using the equation of motion for the
fermion fields, the baryonic contribution to the master function reads as
\begin{eqnarray}
  \E_B &=& 6 \int \frac{d^3 k}{(2 \pi)^3} \, \theta(\mu^{\n}_* -
  e_{\n}(k^\nu)) \frac{(k^x)^2 +
    (m^*_{\n})^2/3}{e_{\n}(k^\nu)}\nonumber \\  
  & +& 6 \int \frac{d^3k}{(2 \pi)^3}  \,\theta(\mu^{\p}_* -
  e_{\p}(k^\nu)) \, J(k,\tilde{k}) \nonumber\\ 
  && \times\frac{(k^x)^2 + (m^*_{\p})^2/3}{\frac{1}{\sqrt{1-v^2}}
    (e_{\p}(k^\nu) + v (k^z - g_\omega \omega^z + g_\rho \rho^z))}
  \nonumber \\ 
  & +& \frac{1}{2} m_\sigma^2 \bar{\sigma}^2 + \frac{1}{2} m_\delta^2
  \bar{\delta}^2 - \frac{1}{2} m_\omega^2 \omega_\mu \omega^\mu -
  \frac{1}{2} m_\rho^2 \rho_\mu \rho^\mu.\ \ \ \ 
\end{eqnarray}
Using the same technique as before, we finally obtain
\begin{eqnarray}
  \E_B &=& \epsilon_{\n}(n_{\n}) + \epsilon_{\p}(n_{\p}) \nonumber \\
  & +& \frac{1}{2} m_\sigma^2 \bar{\sigma}^2 + \frac{1}{2} m_\delta^2
  \bar{\delta}^2 - \frac{1}{2} m_\omega^2 \omega_\mu \omega^\mu -
  \frac{1}{2} m_\rho^2 \rho_\mu \rho^\mu,\ \ \ \  
\end{eqnarray}
where $\epsilon_X(n_X)$ has the form of the energy density of a free
Fermi gas, here computed for the Dirac effective mass of protons and
neutrons, respectively,
\begin{eqnarray}
  \epsilon_X(n_X) =
  \frac{1}{8 \pi^2} &&  \bigg( k_{F,X} \ \mu^X_*
    \,\left(\left(m^*_X\right)^2 + 2 k_{F,X}^2\right) \bigg.\nonumber \\ 
  && \left.- (m^*_X)^4 \ln\left[\frac{k_{F,X} + \mu^X_*}{m^*_X}\right]\right)~.  
\end{eqnarray} 
The quantities $k_{F,X}$ are the Fermi momenta in the respective rest frames, related
to the scalar densities $n_X$ as $k_{F,X} = (3 \pi^2 n_X)^{1/3}$, see (\ref{eq:density}). Electrons can be added trivially at this
point. They are considered as a non-interacting Fermi gas, coupled to
the baryons only via global charge neutrality condition ($n_e= n_{\p}$) such that
finally 
\begin{equation}
  \E = \E_B +\ \epsilon_e(n_{\p})~,
\end{equation}
with $k_{F,e} = k_{F,p}$, $m^*_e = m_e$ and $\mu^e = \sqrt{m_e^2 + k_{F,e}^2}$.

The entrainment matrix is now readily evaluated from the derivatives
of $\E$. To that end, let us first observe that 
\begin{eqnarray}
  m_\omega^2 \omega_\alpha \omega^\alpha &=&
  \frac{g_\omega^2}{m_\omega^2} n_{B}^{\ \alpha}n_{B\ \alpha} \nonumber \\
  &=& -\frac{g_\omega^2}{m_\omega^2} (n_{\n}^2 + n_{\p}^2 + 2 n_{\n}
  n_{\p} \Gamma_\Delta) = - \frac{g_\omega^2}{m_\omega^2} n_B^2,\hspace*{0.8 cm} \\   
  m_\rho^2 \rho_\alpha \rho^\alpha &=& \frac{g_\rho^2}{m_\rho^2}
n_{I}^{\ \alpha}n_{I\ \alpha}  \nonumber \\ 
  &=& -\frac{g_\rho^2}{m_\rho^2} (n_{\n}^2 + n_{\p}^2 - 2 n_{\n} n_{\p}
  \Gamma_\Delta) = -\frac{g_\rho^2}{m_\rho^2} n_I^2.\hspace*{0.8 cm} 
\end{eqnarray}
Secondly, the derivatives of $\E$ with respect to the scalar meson
fields, $\sigma$ and $\delta$, are vanishing by construction; they only
contribute to the Dirac effective masses (\ref{eq:masses}).

As mentioned earlier, within the density-dependent models, the
coupling constants depend on the baryon number density $n_B$ and upon
deriving the entrainment matrix we have to take this dependence into
account, see the definition of the rearrangement term,
Eq.~(\ref{eq:sigmar}). For the derivatives of the master function we
obtain
\begin{subequations}
\begin{eqnarray}
  \mu^{\n} = \frac{\partial\E}{\partial n_{\n}} &=& \mu^{\n}_* + n_{\n}
  a_+ + n_{\p} a_- \Gamma_\Delta + \Sigma^R \frac{\partial n_B}{\partial
    n_{\n}} \label{eq:mus1} \\ 
  \mu^{\p} = \frac{\partial\E}{\partial n_{\p}} &=& \mu^{\p}_* + n_{\p}
  a_+ + n_{\n} a_- \Gamma_\Delta + \Sigma^R \frac{\partial n_B}{\partial
    n_{\p}}  \nonumber\\
    &+& \mu^e \label{eq:mus2}\\ 
  \alpha = \frac{\partial\E}{\partial \Delta^2} &=& \frac{1}{2} n_{\n}
  n_{\p} a_- \Gamma_\Delta^3 + \Sigma^R \frac{\partial n_B}{\partial
    \Delta^2}~, 
\end{eqnarray}
\end{subequations}
In the two-fluid case, $n_B = n_B(n_{\n},n_{\p},\Delta^2)$ and for its
derivatives the following relations hold
\begin{subequations}
\begin{eqnarray}
  \frac{\partial n_B}{\partial n_{\n}} &=& \frac{1}{n_B} ( n_{\n} + n_{\p}
  \Gamma_\Delta)\\ 
\frac{\partial n_B}{\partial n_{\p}} &=& \frac{1}{n_B} (
  n_{\p} + n_{\n} \Gamma_\Delta)\\ 
\frac{\partial n_B}{\partial \Delta^2} &=&
  \frac{1}{2 \,n_B} n_{\n} n_{\p} \Gamma_\Delta^3~.
\end{eqnarray}
\end{subequations}
Using Eqs.~(\ref{K1}) and (\ref{K2}) we finally arrive at the following expressions
for the entrainment matrix
\begin{subequations}
\begin{eqnarray}
  \K^{\n\!\n}  &=& \frac{\mu^{\n}_*}{n_{\n}} + a_+ + \frac{\Sigma^R}{n_B}\\ 
  \K^{\p\!\p}  &=& \frac{\mu^{\p}_*}{n_{\p}} + a_+ +
  \frac{\Sigma^R}{n_B} + \frac{\mu^e}{n_{\p}}\\  
  \K^{\n\!\p}  &=& a_- + \frac{\Sigma^R}{n_B}~.
\end{eqnarray}
\end{subequations}
Different remarks are in order here. First, as can easily be seen, in
the limit of small relative speed, the entrainment matrix elements are
in agreement with the expressions in~\cite{urban2015collective}
derived from Fermi liquid theory to first order in the
velocities\footnote{The elements of the matrix $M$
  in~\cite{urban2015collective} correspond to the $\K^{a b}$
  multiplied by the density of the second index, $M_{ab} = \K^{a b}
  n_b$ (no summation over repeated index).}.  Second,
Eqs.~(\ref{eq:mus1})-(\ref{eq:mus2}) reduce to the chemical potentials
in the single fluid case, see Eqs.~(\ref{eq:mu1})-(\ref{eq:mu2}), in
the limit of vanishing relative speed between both fluids. Finally,
the condition on the entrainment matrix element cited by
\cite{gusakov2009relativistic}, Eq. (8), expressed here as
$u^{\ \alpha}_X p^X_{\ \alpha} = -\mu^X$ is fulfilled (for any
$\Delta^2$), in contrast to the results in
\cite{comer2003relativistic, kheto2014isospin}.

In the numerical implementation, we use the EoS in a tabulated form,
see appendix~\ref{a:num} for more details.  

\subsection[]{Entrainment effects}
\label{sub:entr}

\begin{figure}[!t]
  \center 
  \includegraphics[width = 0.48\textwidth]{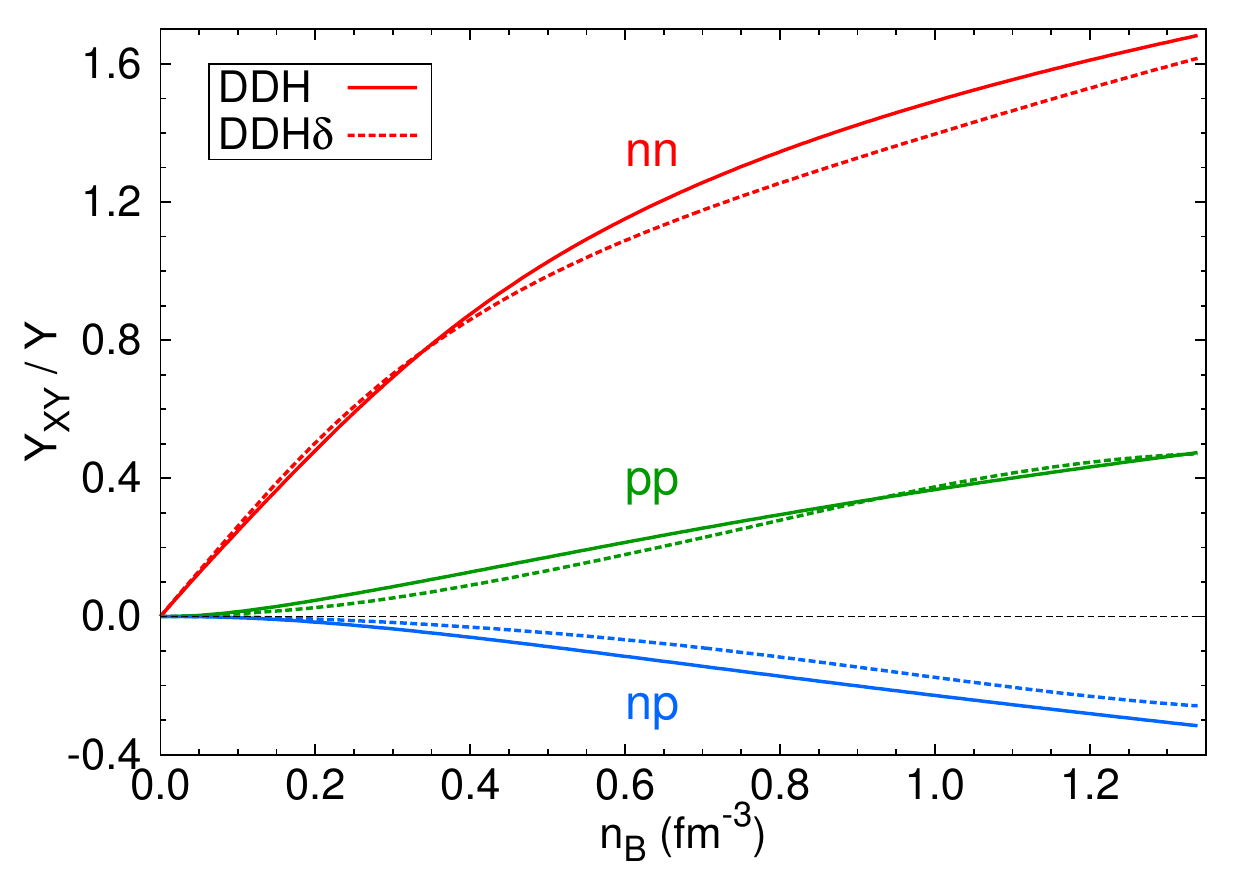}
  \caption{Entrainment coefficients $Y_{X\!Y}$ as functions of total
    baryon density $n_{B}$. Solid (dashed) lines refer to
    DDH($\delta$) EoS. Following the prescription given by Gusakov
    \textit{et al.}~\cite{gusakov2009relativistic}, these coefficients
    are normalized to the constant $Y = 3 n_0 / \mu^{\n}(3n_{0})$,
    where $n_0 = 0.16$ fm$ ^{-3}$ stands for the saturation
    density. For DDH($\delta$), $Y =2.55\times 10^{41}$
    erg$^{-1}$.cm$^{-3}$ ($Y =2.47\times 10^{41}$
    erg$^{-1}$.cm$^{-3}$).}
  \label{Yik}
\end{figure}

Entrainment effects are depicted by the scalar
$\alpha$~(\ref{derthermo2}) which vanishes in the limit where there is
no entrainment. As we do not take the presence of the crust into
account in our model, entrainment is assumed to be only due to the
strong interactions between nucleons. Two different approaches are
commonly followed in the literature to quantify entrainment within the
EoS: either by means of the entrainment matrix coefficients
\citep{gusakov2009relativistic} or by introducing dynamical effective
masses \citep{chamel2006entrainment}.

The elements of the entrainment matrix, ${\mathcal K}^{X\!Y}$, and of
its inverse, $\mathcal{Y}_{X\!Y}$, are functions of three quantities,
e.g. $n_{\n},n_{\p}$ and $\Delta^2$. In order to compare entrainment
effects within different EoSs, it is therefore convenient to study the
limiting case of corotation (\textit{i.e.} $\Delta^2 = 0$) with
$\beta$-equilibrium (see Sec.~\ref{section_structure}). We therefore
introduce the entrainment coefficients~\citep{gusakov2009relativistic}
\begin{equation}
  Y_{X\!Y} \equiv \mathcal{Y}_{X\!Y} |_{\Delta^2 = 0,\ \mu^{\n} = \mu^{\p}}~,
\end{equation}
which depend on a single parameter, e.g. the total baryon density $n_B
= n_{\n} + n_{\p}$.  These coefficients are plotted as functions of
$n_B$ in Fig.~\ref{Yik}, for both EoSs.

\begin{figure*}
  \center 
  \includegraphics[width = 0.48\textwidth]{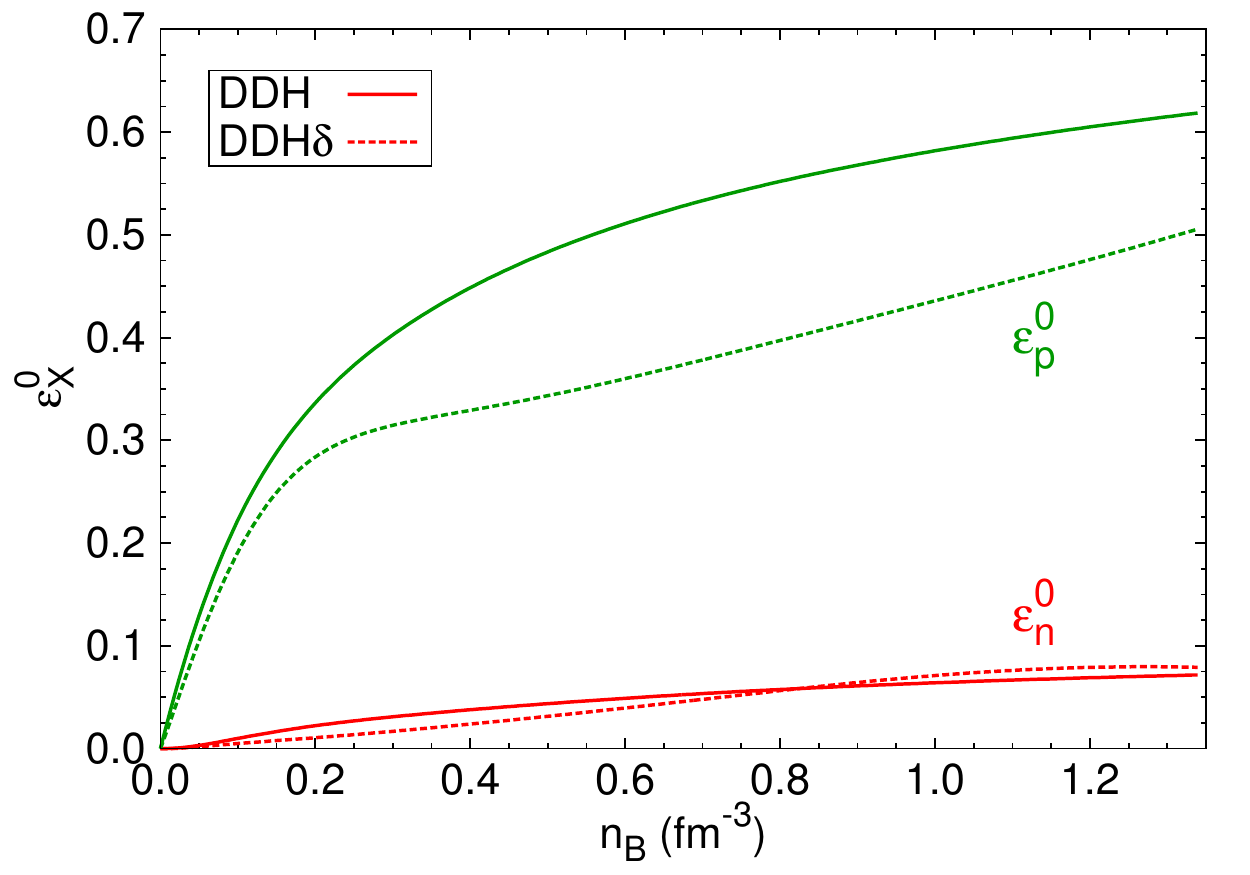}
  \includegraphics[width = 0.48\textwidth]{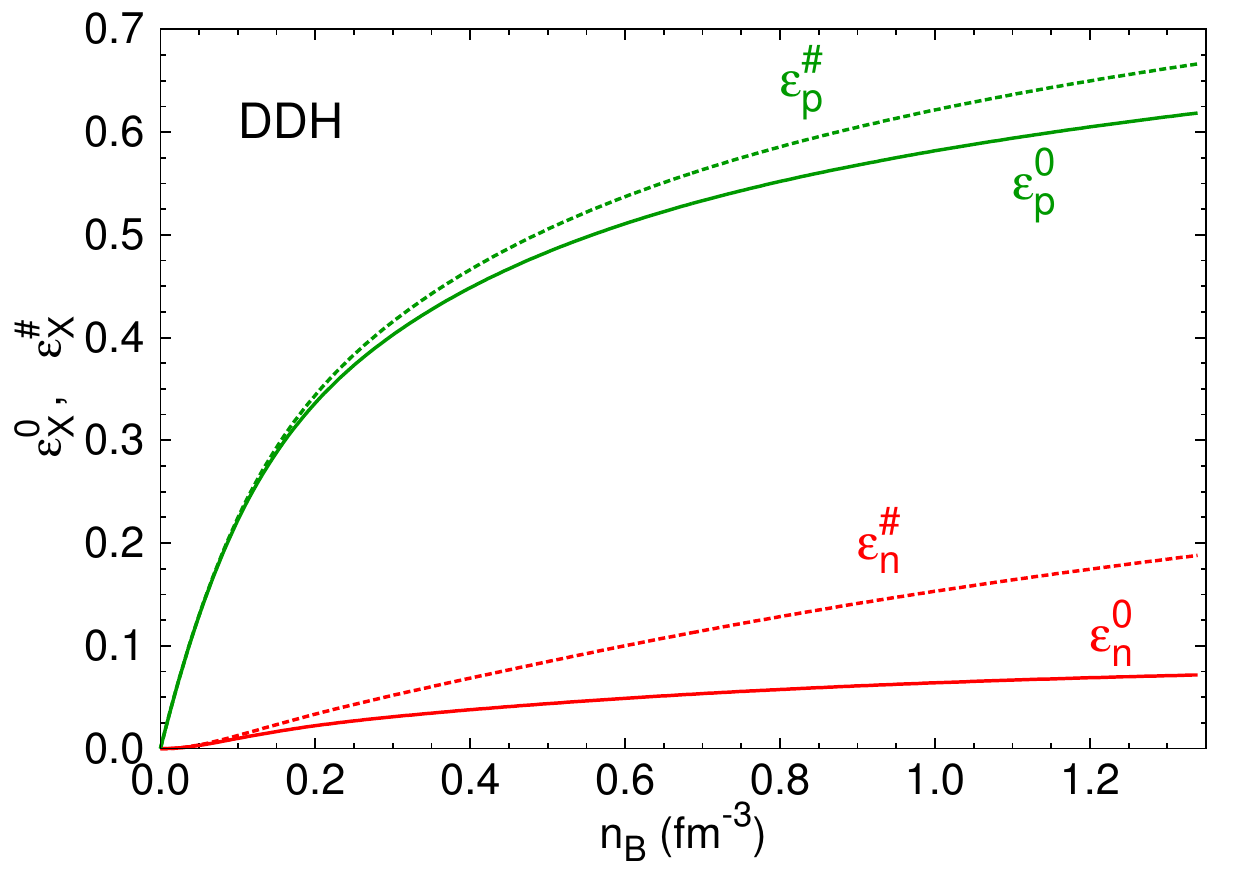}
  \caption{\textbf{Left:} Influence of the interaction on the
    entrainment parameters $\varepsilon_X^0$ . Solid (resp. dashed)
    lines refer to DDH($\delta$) EoS. For better clarity, only
    quantities defined in the zero-velocity frames are
    plotted. \textbf{Right:} Comparison between entrainment parameters
    defined in the zero-velocity frames ($\varepsilon_X^0 $, solid
    lines) and in the zero-momentum frames ($ \varepsilon_X^{\#} $,
    dashed lines), for the DDH EoS.}
  \label{Dyn_mass_comp}
\end{figure*}

In order to study the importance of entrainment effects, we can also introduce
dynamical effective masses. The idea is to describe the dynamics of
each species as if it was alone. Interactions with the other fluid
are included through the effective mass $\tilde{m}_X$ defined as
\begin{equation}
\label{mass_eff}
p_{X}^{\ i} = \tilde{m}_{X}\  u_{X}^{\ i}, 
\end{equation}
where $p_{X}^{\ i}$ and $u_{X}^{\ i}$ stand for the
spatial parts of the conjugate momentum and the 4-velocity of fluid
$X$, respectively. Such a definition is formulated in the rest-frame of the
background, \textit{i.e.} the second fluid $Y$.

As already noticed by Prix \textit{et al.}~\cite{prix2002slowly}, it
is not possible to define the rest frame for fluid $Y$ in a unique
way.  In the \textit{zero-velocity} frame of the fluid $Y$, where
$u_{Y}^{\ i}= (0, 0, 0)$, Eq. (\ref{matrice}) becomes
\begin{equation}
 p_{X}^{\ i} = \mathcal{K}^{X\! X} n_X\   u_{X}^{\ i},
\end{equation}
such that, using (\ref{K1}), we obtain 
\begin{equation}
\tilde{m}_{X} = \mathcal{K}^{X\! X} n_X = \mu^X \left(1 - \varepsilon_X  \right), 
\end{equation}
where we have introduced the quantity 
\begin{equation}
\label{def:epsilon}
\varepsilon_X = \frac{2 \alpha}{ n_X \mu^X \Gamma_{\Delta}^2 }.
\end{equation}
Assuming again corotation\footnote{In the corotating limit, one should
  notice that $u_{Y}^{\ i}=\!  u_{X}^{\ i}=0$, so that it is not
  possible to define an effective mass from (\ref{mass_eff}). Strictly
  speaking, the quantity $m^0_X$ has no real physical meaning but is
  convenient to compare different EoSs. Note that, the relative speed
  in neutron stars being very small, $\tilde{m}_X\simeq
  m^0_X$. Similar remarks apply to $m_X^{\#}$.} and
$\beta$-equilibrium, the following effective mass can be
introduced~\citep{prix2002slowly, chamel2006entrainment}
\begin{equation}
\label{mass_star}
m^0_X \equiv \K^{XX}|_{\Delta^2 = 0,\ \mu^{\n} = \mu^{\p}}\  n_X =
\mu^X \left(1 - \varepsilon_X^0 \right),
\end{equation}
where $\varepsilon^0_X = \varepsilon_X|_{\Delta^2 = 0,\ \mu^{\n} =
  \mu^{\p}}$. In the \textit{zero-momentum} frame of the fluid $Y$,
where $p_{Y}^{\ i}= (0, 0, 0)$, Eq. (\ref{matrice}) leads to
\begin{equation}
 p_{X}^{\ i} = \frac{n_X}{\mathcal{Y}_{X\! X}}\   u_{X}^{\ i},
\end{equation}
from which, we obtain
\begin{equation}
  \tilde{m}_{X} = \frac{n_X}{\mathcal{Y}_{X\! X}} = \mu^X \left(1 -
    \varepsilon_X \frac{1 -   \varepsilon_Y \Delta^2}{1 -
      \varepsilon_Y}\right).  
\end{equation}

This leads us to introduce another effective mass for fluid
$X$~\citep{prix2002slowly, chamel2006entrainment} 
\begin{equation}
  \label{mass_shar}
  m^{\#}_X  \equiv   \frac{n_X}{Y_{XX}} = \mu^X \left(1 - \varepsilon_X^{\#} \right), 
\end{equation}
where 
\begin{equation}
  \varepsilon_X^{\#}= \frac{\varepsilon^0_X}{1-\varepsilon^0_Y}.
\end{equation} 

The quantities $m_{\n}^0$ and $m_{\n}^{\#}$ introduced so far are
linked to the quantities $\varepsilon_ {\text{mom}}$ and $\varepsilon_
{\text{vel}}$ studied by \cite{comer2003relativistic} through
\begin{equation}
m_{\n}^0 = \varepsilon_ {\text{vel}} m_{\n} \ \ \text{and} \ \
m_{\n}^{\#} = \frac{ 1}{\varepsilon_ {\text{mom}}}m_{\n}. 
\end{equation}

The entrainment parameters $\varepsilon^0_X$ are shown in
Fig.~\ref{Dyn_mass_comp}~-~left, for the two different EoSs, as
functions of the total baryon density. We do not show the dynamical
masses since they contain not only entrainment effects, but also
(special) relativistic corrections. In fact, for vanishing
entrainment, \textit{i.e.} $\alpha = 0$, the effective masses
(\ref{mass_star}) and (\ref{mass_shar}) reduce to the chemical
potentials (since all forms of energy contribute to the mass), not the
bare masses. The parameters $\varepsilon_X$ are, on the contrary, a
good measure of the importance of entrainment effects.

As can be seen in Fig.~\ref{Dyn_mass_comp}, entrainment effects become
more and more important as the baryon density increases, where the
interaction between particles gets stronger. Entrainment effects are
quite important on the proton fluid beyond saturation density, whereas
the neutron fluid is much less affected. This is simply a consequence
of the relative proportion of the two fluids, $\varepsilon_{\n}
=\frac{n_{\p}}{n_{\n}}\varepsilon_{p}$, when $\beta$-equilibrium is
enforced. Note that we checked that the stability conditions derived
by \cite{chamel2006entrainment}, \textit{i.e.}
\begin{equation}
  0\leq \varepsilon^0_{\n} < x_{\p} \ \ \text{and} \ \ 0\leq
  \varepsilon^0_{\p} < 1 - x_{\p}, 
\end{equation}
where $x_{\p} = n_{\p}/n_B$, are verified. Results in the
zero-momentum frame are very similar (see Fig.~\ref{Dyn_mass_comp} -
right), except at very high densities. The neutron fluid, anyway, is
much less affected by entrainment and both parameters remain small
with neutron effective masses close to $\mu^{\n}$. Comparing both
EoSs, the general behavior is very similar. The discrepancy on the
proton entrainment within the two EoSs, that is visible at high $n_B$,
is due to the very different proton ratios (at $\beta$-equilibrium)
predicted by these EoSs at a given $n_B$, as a consequence of the
different values of symmetry energies and slopes at saturation density
(see Table~\ref{tab:nuclear}). As the neutrons are much more numerous,
the influence of the different proton ratios on the neutron
entrainment is smaller.

\begin{figure*}
\includegraphics[width = 0.49\textwidth]{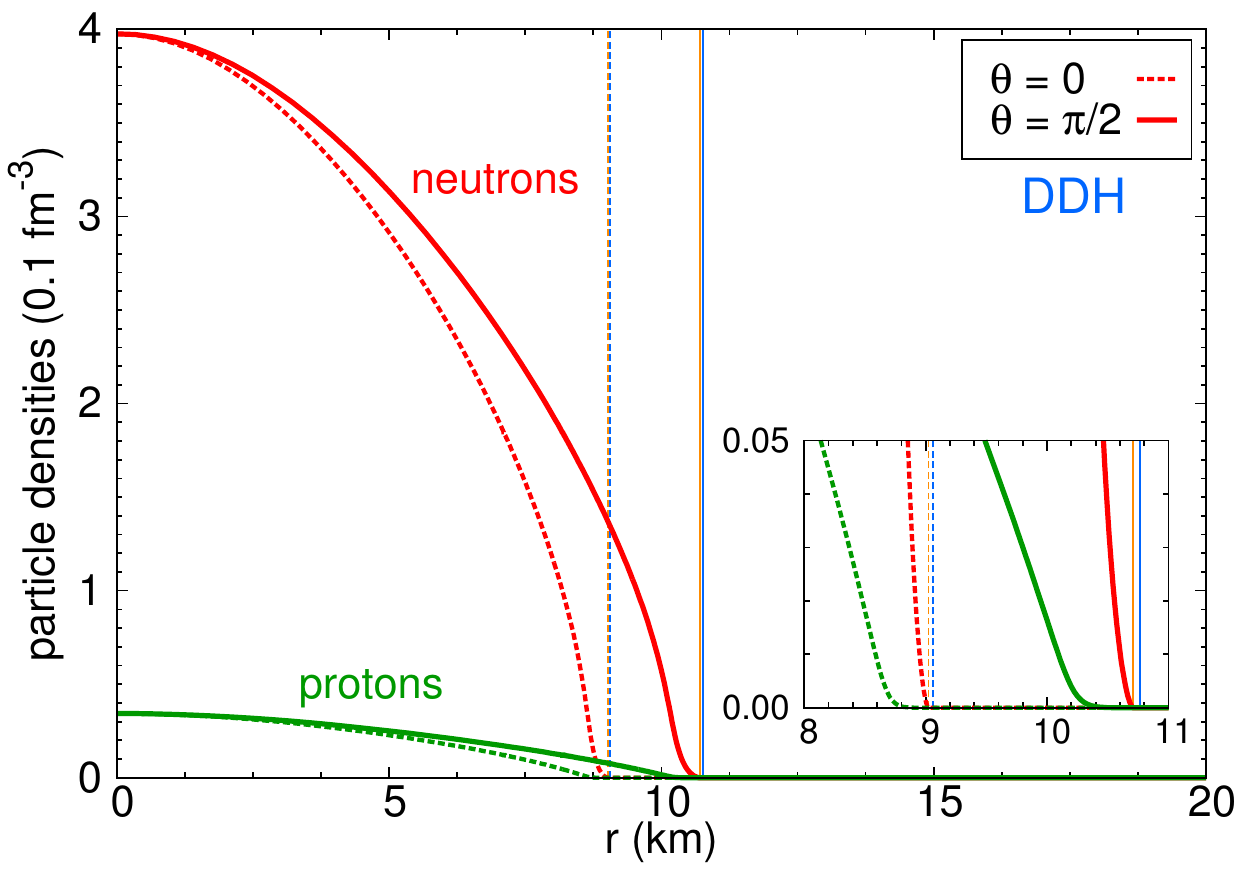}
\includegraphics[width = 0.49\textwidth]{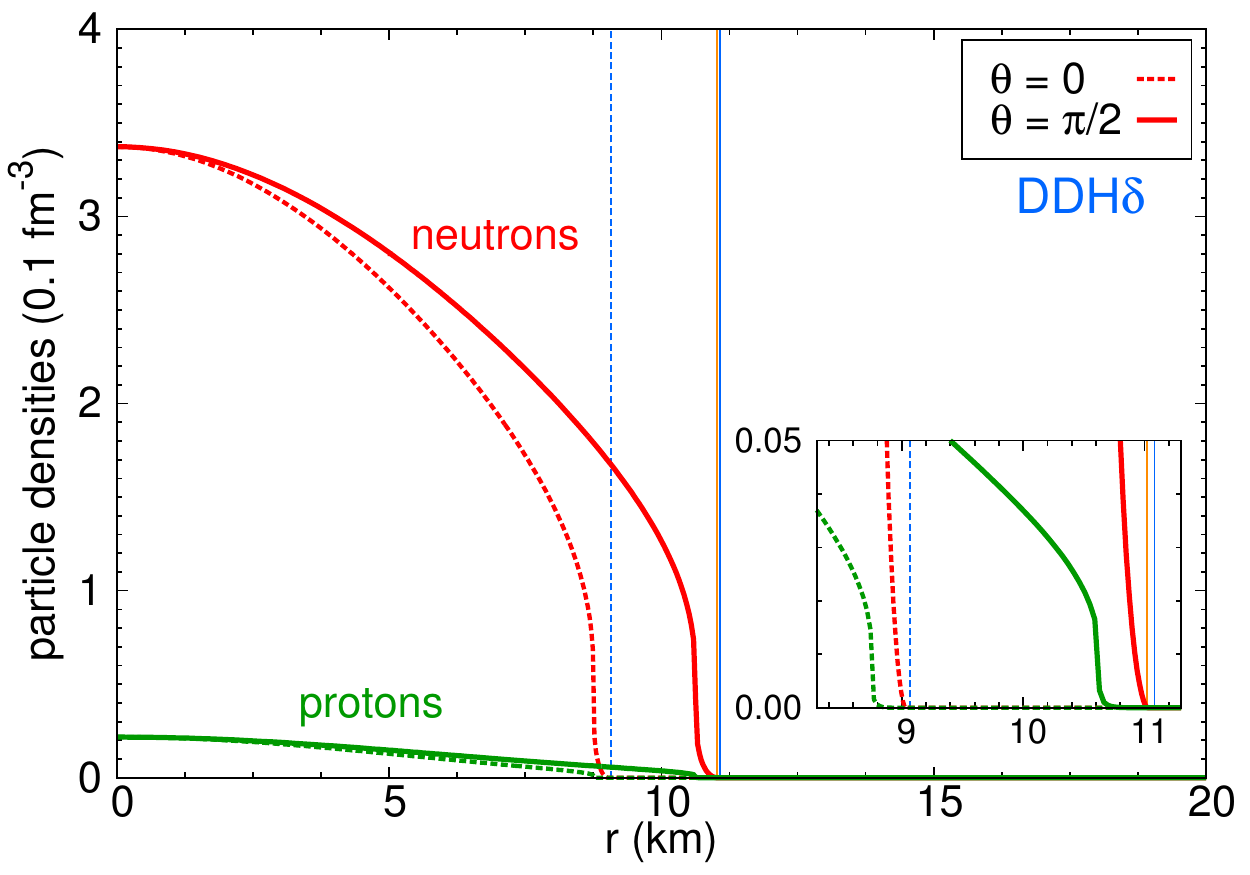}
\caption{Density profiles $n_{\n}$ and $n_{\p}$ plotted with respect to the radial
  coordinate $r$ for a star with $M_{G} = 1.4$ M$_{\odot}$
  spinning at $\Omega_{\n}/2\pi = \Omega_{\p}/2\pi = 716$ Hz, obtained
  with DDH and DDH$\delta$ EoSs. Neutron (protons) particle density
  $n_{\n}$ ($n_{\p}$) is plotted in red (green). Dashed and solid
  lines refer to profiles obtained in the polar and equatorial
  planes. Blue and orange vertical lines represent the coordinate of
  vanishing density of protons and neutrons. Some zooms of the area
  surrounding the surfaces are also presented. }
  \label{densities}
\end{figure*}

\section[]{Equilibrium configurations}
\label{eqconf}

We now use the model described in the previous sections to get some
realistic equilibrium configurations describing superfluid neutron
stars. Some general results were already discussed in Prix \textit{et
  al.}~\cite{prix2005relativistic}. Here, we mainly focus on the
consequences of taking realistic EoSs into account.

For the different configurations studied in this Section, the virial
identity violations
\citep{gourgoulhon1994formulation,bonazzola1994virial}, which are
useful checks of the accuracy of numerical solutions of Einstein
equations, are of the order of $\sim 10^{-8}-10^{-5}$ depending on the
mass of the neutron star, the rotation rates and the choice of the
grid used to describe the star. This means that the numerical errors
in our models should be below this value and gives us confidence in
the accuracy of our results.

\subsection[]{Density profiles}

Assuming corotation and $\beta$-equilibrium, the external fluid
appears to be always the proton fluid, because
$m_{\p}~\lesssim~m_{\n}$. A more realistic model would consider the
presence of an elastic crust below the surface of the star.  For the
DDH$\delta$ EoS, the maximum mass predicted is $2.16$ M$_{\odot}$ in
the static case and increases up to $2.21$ M$_{\odot}$ for 716 Hz, the
highest rotation frequency observed today~\cite{hessels2006radio}. The
maximum masses obtained with the DDH EoS are a bit smaller: $2.08$
M$_{\odot}$ for static configurations and $2.12$ M$_{\odot}$ at 716
Hz.  These values are consistent with the accurate measurements of
2~M$_{\odot}$ neutron stars in binary pulsars \citep{demorest2010two,
  antoniadis2013massive}. We refrain from giving radius values here,
since our model does not contain any elastic crust, inducing an error
of the order 500 m in the radius determination.

Keeping $\beta$-equilibrium at the center of the star, and allowing
for a relative lag of up to $\left(\Omega_{\n} - \Omega_{\p}\right) /
\Omega_{\p} \sim 1.4 \times 10^{-3}$, the relative increase of the
maximum mass is $\sim 6 \times 10^{-5}$. Such a lag is well beyond the
maximum lag expected in neutron stars from the glitch amplitude (see
footnote in Sec.~\ref{section_structure}). We thus conclude that the
maximum mass is very precisely determined in the corotation
approximation.

Assuming again corotation and $\beta$-equilibrium, we plot the density
profiles obtained from the two EoSs as functions of the radial
coordinate $r$ in Fig. \ref{densities} for a star whose gravitational
mass is $ 1.4$ M$_{\odot}$, with a rotation rate $\Omega_{\n}/2\pi =
\Omega_{\p}/2\pi = 716$ Hz. Profiles in the equatorial (polar) planes
are shown in solid (dashed) lines. It can be nicely seen in the zoom
(right panel) that the proton fluid is the external fluid.  As
expected, protons are much less abundant than neutrons. At the center
of the star ($r=0$), the proton ratio is $x_{\p} = n_{\p}/n_B \simeq
0.08$ for the DDH EoS, whereas $x_{\p}(r=0)\simeq 0.06$ for the
DDH$\delta$ EoS. Using the DDH EoS, the central baryon density is
equal to $n_B(r=0)\simeq 0.44$ fm$^{-3}$, which is close to three
times the saturation density. With the DDH$\delta$ EoS, it is smaller,
$n_B(r=0)\simeq 0.36$ fm$^{-3}$. The difference comes from the fact
that for $\beta$-equilibrated matter at a given $n_B$ relevant for
neutron stars, as can be inferred from symmetry energy and slope the
pressure is systematically higher in DDH$\delta$ than in
DDH. Therefore, for the same gravitational mass of the star, $n_B$ is
lower. Here we do not study the influence of a difference in rotation
rates between both fluids because from the astrophysical side it is
expected to be so small that the results would be very similar to
those presented in Fig.~\ref{densities} and, on the other hand, many
results concerning models with arbitrarily different rotation rates
were presented in Prix \textit{et al.}~\cite{prix2005relativistic}.

\subsection[]{Angular momenta}

\begin{figure}
  \center 
  \includegraphics[width = 0.48\textwidth]{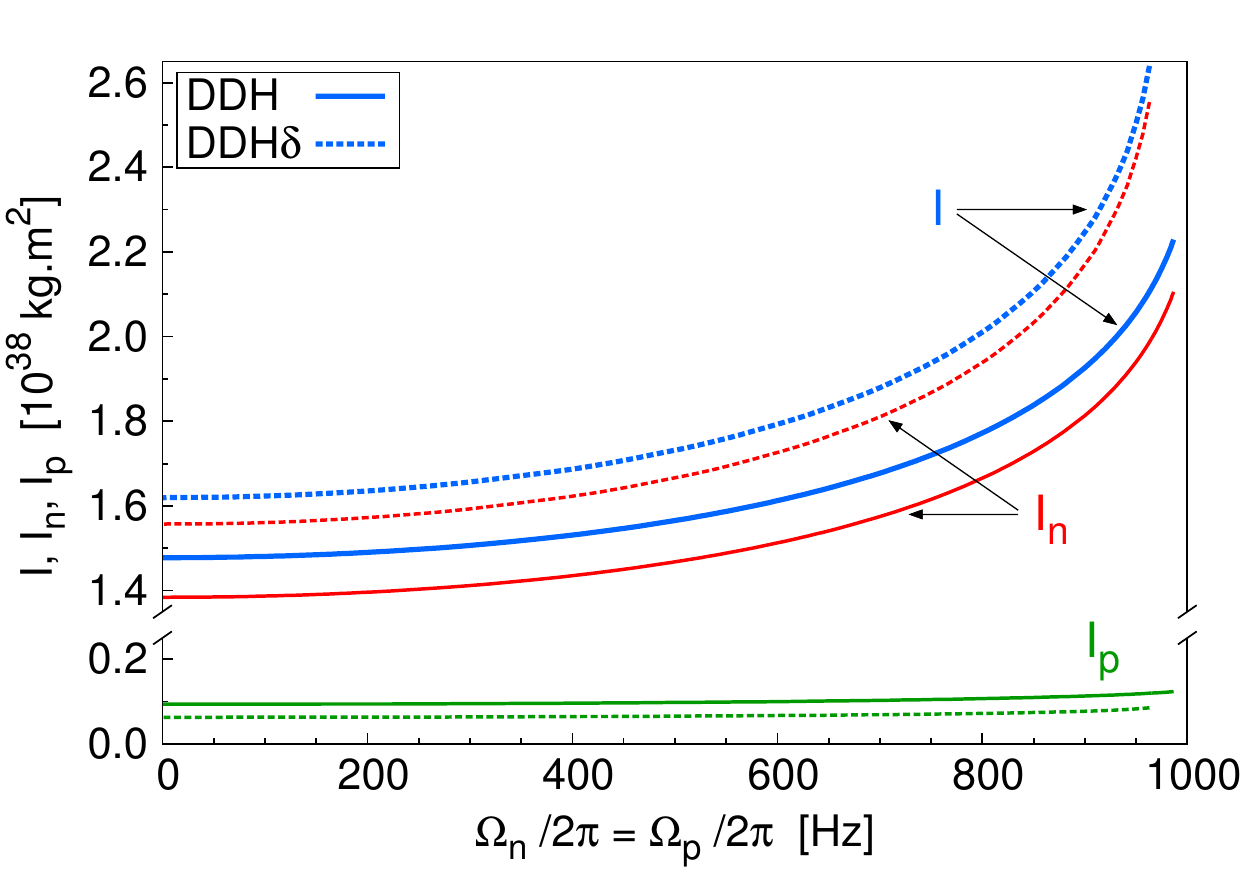}
  \caption{Moments of inertia plotted with respect to the angular
    velocity of the pulsar, taking both angular velocities to be equal
    and assuming $\beta$-equilibrium, for a (same) fixed total baryon
    mass $M^B=1.542$ M$_{\odot}$ corresponding to
    $M_G \simeq 1.4$ M$_{\odot}$. Results from the DDH
    (DDH$\delta$) EoS are represented with solid (dashed) lines. Total
    quantities are shown in blue whereas neutron (proton) ones are
    plotted in red (green).}
  \label{JandI}
\end{figure}

We give here some results on the angular momenta, as well as for
moments of inertia defined in Sec.~\ref{global_quantities}.  The
moments of inertia $I$, $I_{\n}$ and $I_{\p}$ are plotted as functions
of the angular velocity of the star in Fig. \ref{JandI}, assuming
$\Omega_{\n} = \Omega_{\p}$. Here is considered a sequence with
constant total baryon mass, corresponding to neutron stars whose
gravitational masses are around $1.4$ M$_{\odot}$. At low angular
velocities, the moments of inertia are nearly constant, such that the
angular momenta depend linearly on $\Omega_{X}$. Approaching Keplerian
velocity, this is no longer the case and momenta of inertia and
angular momenta are steeply increasing. As expected, the total angular
momentum of the star is dominated by the neutron angular momentum.

\begin{figure}
  \center 
  \includegraphics[width = 0.48\textwidth]{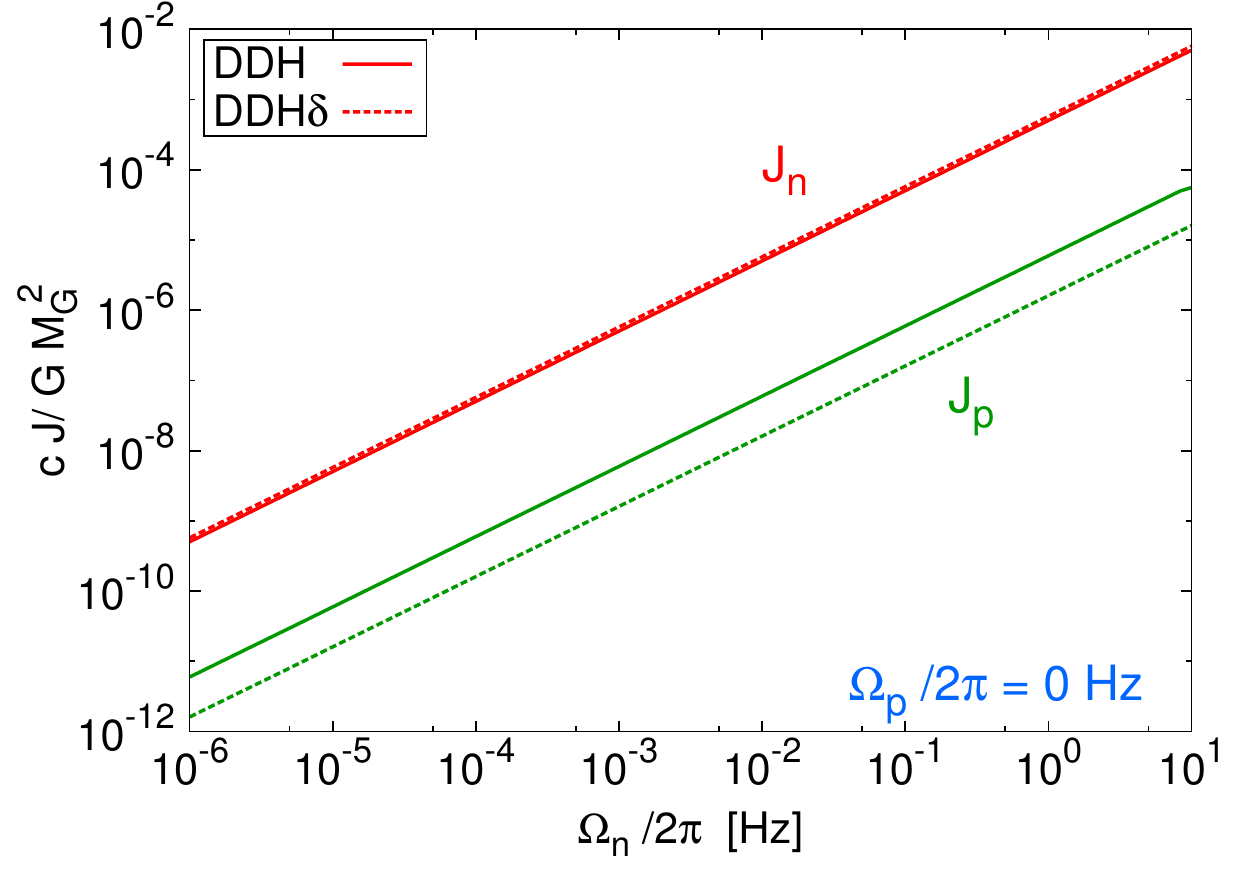}
  \caption{Neutron and proton angular momenta as functions of the
    neutron angular velocity $\Omega_{\n}$, assuming the proton fluid
    to be at rest with respect to a static observer at spatial infinity
    ($\Omega_{\p}/2\pi = 0 $ Hz). Results from the DDH (DDH$\delta$)
    EoS are shown with solid (dashed) lines.  Quantities are plotted
    for a fixed total baryon mass (equal for the two EoSs), assuming $\beta$-equilibrium at the center. }
  \label{J_entra}
\end{figure}

Note that in the present two-fluid case the angular momentum of a
fluid can be nonzero even if its angular velocity vanishes. Two
different effects can be identified at the origin of this
phenomenon. The first one is the general relativistic frame-dragging
effect, which can be seen through the presence of the metric term
$\omega$ in the definition of the physical velocities
(cf. Eq. (\ref{Un})). Although $\Omega_{\p} = 0$, the rotation of the
neutron fluid ($\Omega_{\n} \neq 0$) thus leads to $U_{\p} <0$ and to
a non-vanishing proton angular momentum (see (\ref{def1}) and
(\ref{def2})). The second contribution refers to the dependence of the
proton angular momentum on the physical velocity of the neutron fluid
as a consequence of entrainment (see e.g. (\ref{explicite})), which is
clearly visible on Eq. (\ref{newto}) in the Newtonian limit. To
illustrate this phenomenon, in Fig. \ref{J_entra} two sequences of
stars (corresponding to the two EoSs) are plotted as a function of
$\Omega_{\n}$, assuming $\Omega_{\p} = 0 $ Hz, for a fixed baryon
mass. Although the proton angular velocity vanishes, its angular
momentum is nonzero, rising roughly linearly with
$\Omega_{\n}$. Entrainment gives thereby the dominant contribution,
since $J_{\p}$ is positive, but the frame-dragging effect (which acts
on the proton angular momentum in an opposite way to entrainment)
contributes almost as much as entrainment.

\section[]{Conclusion}
Both microscopic calculations and observations give strong indications
that the interior of neutrons stars contains superfluid matter.
Superfluidity is thus an important ingredient that needs to be taken
into account in order to build realistic models for neutron stars,
which could be very useful for the study of oscillations, glitches and
cooling phenomena.

As a first step towards this objective, we have extended the numerical
model of stationary rotating superfluid neutron stars proposed by Prix
\textit{et al.}~\cite{prix2005relativistic} to the use of realistic
EoS. These models consider the neutron star to be composed of two
fluids, neutrons and charged particles (protons and electrons), which
are free to rotate uniformly around a common axis with different
angular velocities. Obviously, these models can be applied for any
rotation frequency and go therefore beyond the slow rotation
approximation models of Refs.~\cite{andersson2001slowly,
  comer2004slowly, kheto2015slowly}. To reach high numerical accuracy,
tabulated two-fluid EoSs were interpolated with a high-order
thermodynamically consistent scheme, that we tested on analytic
EoSs. An overall precision of $10^{-7}$-$10^{-8}$, measured via
violations of the virial theorem, could be reached. This is of the
same order as typical one-fluid models employing realistic EoS. These
are first numerical model of rapidly rotating neutron stars in full
general relativity and with realistic EoSs.

For these numerical models we need the EoS depending on the two
densities and the relative velocity, i.e. $\E(n_n, n_p,
\Delta^2)$. To this end, following the spirit of Comer and
Joynt~\cite{comer2003relativistic}, we have presented a formalism to
calculate the EoS at an arbitrary value of $\Delta^2$. Entrainment
parameters have been derived from this EoS. We have shown that in the
limit of small $\Delta^2$ our entrainment parameters are in agreement
with those derived from Fermi liquid theory to lowest order in the
relative velocities. This means that the large numerical differences
between the entrainment parameters calculated on the one hand in
Refs.~\cite{comer2003relativistic,kheto2014isospin,kheto2015slowly}
from the EoS and on the other hand in
Refs.~\cite{gusakov2009relativistic,gusakov2014physics} from
Fermi liquid theory simply stem from the fact that the relativistic
deformation of the Fermi spheres has not been taken into account in
the calculations of
Refs.~\cite{comer2003relativistic,kheto2014isospin,kheto2015slowly}.
We have applied our new formalism to two density-dependent RMF
parametrizations, DDH and DDH$\delta$, being consistent with standard
nuclear matter and neutron star properties. The entrainment parameters
are qualitatively very similar in both models. If $\beta$-equilibrium
is imposed, entrainment has a stronger effect on the proton fluid due
to the low proton fraction. Quantitatively, the difference between
both models is non-negligible only for the proton fluid, the higher
proton fraction in DDH leading to more pronounced entrainment effects
than in DDH$\delta$.

As a first application, we have computed several relativistic
equilibrium configurations. As expected, maximum masses are only
marginally influenced by entrainment and a small lag in rotation
frequencies of the two fluids. We did not discuss radii since our
models do not contain any crust, and the extracted radii would thus
not be reliable. Entrainment is more important for the determination
of angular momenta and moment of inertia. In particular, the angular
momentum of one fluid can be nonzero even if its angular velocity is
vanishing. The entrainment induces thereby an opposite effect to
relativistic frame dragging. We have shown that with our EoS,
entrainment is slightly more important than frame-dragging, leading to
a positive angular momentum for the non-rotating fluid.

In this paper, we mainly focused on the properties of neutron stars
cores assuming homogeneous matter. As already mentioned before,
entrainment effects are expected to be much stronger in the solid
crust due to Bragg scattering of dripped neutrons off nuclear
clusters~\cite{chamel2005band,chamel2012neutron}. An interesting
extension of this work would thus be to include the presence of a
solid crust.  We also plan to use the models discussed here for the
study of quasi-stationary evolution of neutron stars, as could be
found during glitches.

\begin{acknowledgments}
We would like to thank Nicolas Chamel for instructive discussions and
Elena Kantor for useful comments. This work has been partially funded
by the SN2NS project ANR-10-BLAN-0503, the ``Gravitation et physique
fondamentale" action of the Observatoire de Paris, and the COST action
MP1304 ̀̀``NewComsptar".
\end{acknowledgments}

\appendix

\section{Newtonian limit of the angular momenta}
\label{Jnewt}
Here, we study the Newtonian limit of Eqs. (\ref{def1}) and (\ref{def2}). To do so, we rewrite the two angular momentum densities as 
\begin{equation}
\label{explicite}
\left\{
  \begin{array}{rcl}
    j^{\n}_{\varphi} &=& \left[ n_{\n} \Gamma_{\n}^2 \mu^{\n} \ U_{\n}    +
      2 \alpha \frac{\Gamma_{\n}^2}{\Gamma_{\Delta}^2}
      \left(\frac{\Gamma_{\p}}{\Gamma_{\Delta}\Gamma_{\n}} U_{\p} -
        U_{\n} \right)\right]  B r \sin \theta, \\[+5 pt]	 
    j^{\p}_{\varphi} &=& \left[ n_{\p} \Gamma_{\p}^2 \mu^{\p} \ U_{\p}     +
      2 \alpha
      \frac{\Gamma_{\p}^2}{\Gamma_{\Delta}^2}
      \left(\frac{\Gamma_{\n}}{\Gamma_{\Delta}\Gamma_{\p}} U_{\n} -
        U_{\p} \right) \right] B r \sin \theta.  
  \end{array}
\right.
 \end{equation}

 In the Newtonian limit, the different quantities appearing in
 Eq. (\ref{explicite}) simplify as $\Gamma_{\n} \simeq \Gamma_{\p}
 \simeq \Gamma_{\Delta}\simeq 1$, $\mu^{\n} \simeq m_{\n}$ and
 $\mu^{\p} \simeq m_{\p}$. Thus, Eq. (\ref{explicite}) becomes
\begin{equation}
\label{explicite2}
\left\{
  \begin{array}{rcl}
	j^{\n}_{\varphi} &=& \left[n_{\n} m_{\n} \ U_{\n}    +
	2 \alpha \left( U_{\p} - U_{\n} \right)\right] r \sin \theta, \\[3 pt]
	j^{\p}_{\varphi} &=& \left[ n_{\p}m_{\p} \ U_{\p}     +
	2 \alpha \left(U_{\n} - U_{\p} \right) \right] r \sin \theta,
  \end{array}
\right.
 \end{equation}
where the physical velocities verify
 \begin{equation}
 \label{veloNewt}
 U_{\n} \simeq \Omega_{\n} r\sin \theta \ \ \text{and} \ \ U_{\p}
 \simeq \Omega_{\p} r\sin \theta. 
 \end{equation}

 Considering that $A \rightarrow 1$ and $B \rightarrow 1$, the element
 volume $\df^{\, 3}\! \Sigma$ tends towards $\df^{\, 3}\!
 \Sigma_{\text{f}} = r^2\sin \theta \df r \df \theta \df \varphi$,
 which is the element volume of the flat spacetime. Replacing
 Eq. (\ref{explicite2}) in Eq. (\ref{def2}), the non relativistic
 limit of the angular momentum of the two fluids reads as
 \begin{equation}
\label{def3}
\left\{
  \begin{array}{rcl}
	J_{\n} &=& \int_{\Sigma_{t}}n_{\n}m_{\n} \left( \Omega_{\n} +
          \varepsilon_{\n} \left(\Omega_{\p} - \Omega_{\n} \right)
        \right) r^2 \sin^2 \theta \df^{\, 3}\! \Sigma_{\text{f}} ,
        \\[3 pt] 
	J_{\p} &=& \int_{\Sigma_{t}}n_{\p}m_{\p} \left( \Omega_{\p} +
          \varepsilon_{\p} \left(\Omega_{\n} - \Omega_{\p} \right)
        \right) r^2 \sin^2 \theta \df^{\, 3}\! \Sigma_{\text{f}} ,   
  \end{array}
\right.
\end{equation} 
where the entrainment parameters $\varepsilon_{\n}$ and
$\varepsilon_{\p}$ are defined as
\begin{equation}
\label{eps_newt}
\varepsilon_{\n} n_{\n} m_{\n} = 2 \alpha = \varepsilon_{\p} n_{\p} m_{\p},
\end{equation}
see Eq.~(\ref{def:epsilon}). Assuming the two angular velocities to be
uniform and introducing the moment of inertia of fluid $X$  
\begin{equation}
\label{I_newt}
	I_{X} = \displaystyle \int_{\Sigma_{t}} n_{X} m_{X}  r^2 \sin^2\theta \df^{\, 3}\! \Sigma_{\text{f}},
 \end{equation}
and its corresponding mean coupling term
\begin{equation}
\tilde{\varepsilon}_X = \frac{\displaystyle \int_{\Sigma_{t}}
  \varepsilon_X n_{X} m_X  r^2 \sin^2\theta \df^{\, 3}\!
  \Sigma_{\text{f}}}{\displaystyle \int_{\Sigma_{t}}  n_{X} m_X  r^2
  \sin^2\theta \df^{\, 3}\! \Sigma_{\text{f}}},  
 \end{equation}
 
 the two Newtonian angular momenta read as 
 \begin{equation}
 \label{newto}
 \left\{
   \begin{array}{rcl}
J_{\n}  &=& I_{\n} \Omega_{\n} + I_{\n} \tilde{\varepsilon}_{\n}
\left(\Omega_{\p} - \Omega_{\n} \right), \\[3 pt] 
J_{\p}  &=& I_{\p} \Omega_{\p} + I_{\p} \tilde{\varepsilon}_{\p}
\left(\Omega_{\n} - \Omega_{\p} \right), 
  \end{array}
\right. 
 \end{equation}
in agreement with the results by Sidery \textit{et al.}~\cite{sidery2010dynamics}.

\section{Numerical implementation of the tabulated EoS}\label{a:num}


Considering a tabulated EoS leads to two additional kinds of numerical
errors, linked to the accuracy with which the table is computed and
the precision of the interpolation scheme.

\begin{figure}[t]
\center 
\includegraphics[width = 0.4\textwidth]{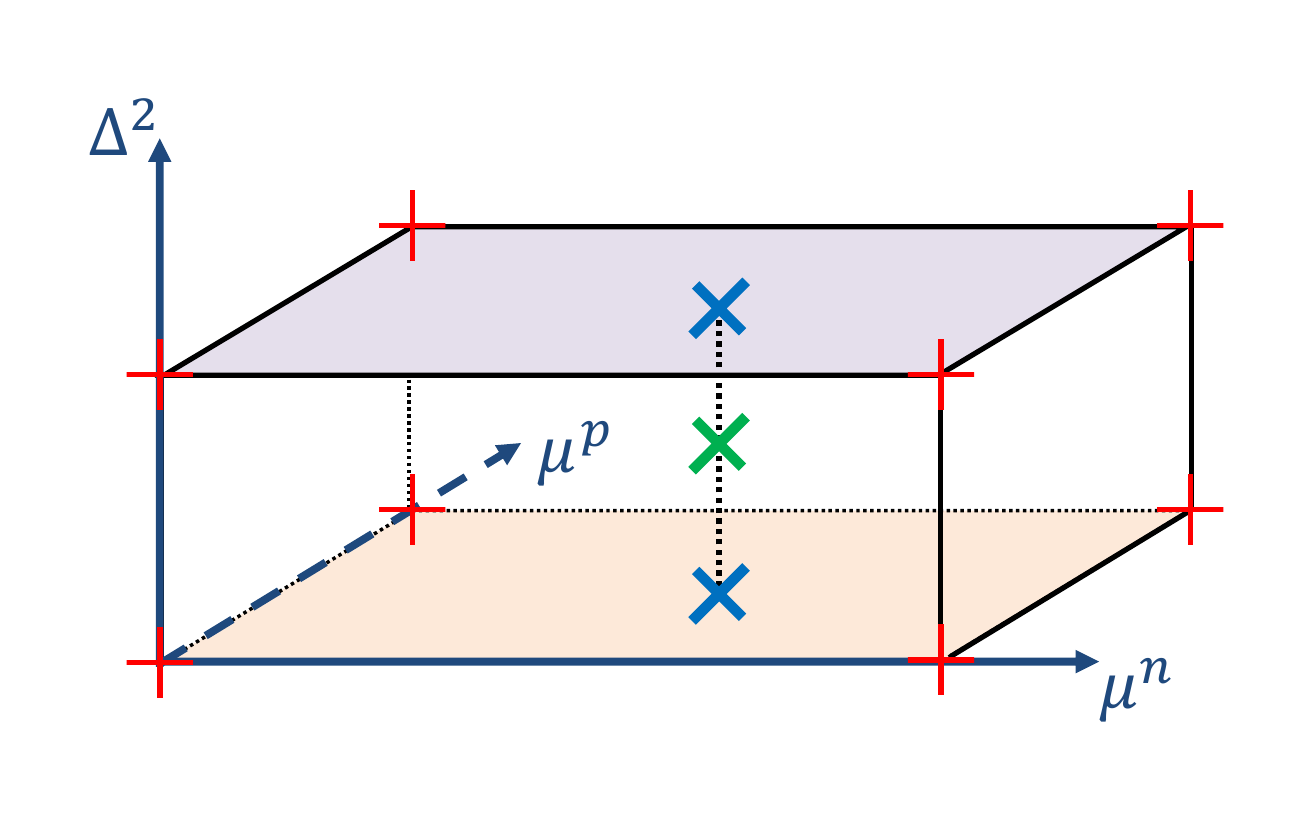}
\caption{3D interpolation scheme on a parallelepipedic grid (red
  crosses), in a point corresponding to a given value of $(\Delta^2,
  \mu^{\n},\mu^{\p})$ (green cross). On each plan where $\Delta^2$ is
  constant, quantities are interpolated with a 2D thermodynamically
  consistent method on the chemical potentials (blue crosses). From
  these two values, a linear interpolation is used in $\Delta^2$, in
  order to obtain the values of the quantities needed at the
  interesting point.}
\label{schemainterpol}
\end{figure}

For each iteration step in the numerical procedure, the matter sources
involved in the Einstein equations are computed from the values of
$H^{\n}$, $H^{\p}$ and $\Delta^2$ at every grid points (see
\cite{prix2005relativistic}). We then use the EoS in the form of the
pressure $\Psi(\mu^{\n},\mu^{\p}, \Delta^2)$
(cf. Eq. (\ref{thermo2})), instead of the energy density $\E$. For
each EoS, we build a table using a grid made of parallelepipeds in the
relative speed $\Delta^2$ and the chemical potentials $ \mu^{\n}$ and
$\mu^{\p}$ (see Fig. \ref{schemainterpol}), which contains, for a
given value of this triplet, the set of variables required to the
interpolation. As the different thermodynamic quantities can be
expressed as functions of the interpolated values of $\Psi$, $n_{\n}$,
$n_{\p}$ and $\alpha$ (cf. Eqs. (\ref{K1}) and (\ref{K2})), we need a
scheme able to interpolate with high precision a function and its
first derivatives (cf. Eqs. (\ref{derthermo1}) and
(\ref{derthermo2})).

To do so, we use the thermodynamically consistent interpolation based
on Hermit polynomials presented by
\cite{swesty1996thermodynamically}. Unfortunately, one can not
directly employ this high-order method on the triplet
$(\mu^{\n},\mu^{\p}, \Delta^2)$, because it would require the presence
of 3-order derivatives in the table, which are extremely difficult to
compute with sufficient precision. Instead, the 3D interpolation
scheme we implemented is the following (see Fig.\ref{schemainterpol}):
\begin{enumerate}[font= \bfseries \color{blue},  align = right, leftmargin=*]
\item One starts by locating in the table the triplet $\left(\Delta^2,
    \mu^{\n}, \mu^{\p}\right)$ in which the interpolation is required, 
\item On the two planes with constant $\Delta^2$ surrounding this
  point, we carry out a 2D thermodynamically consistent interpolation
  in the chemical potentials on $\Psi$ (which also gives the values of
  $n_{\n}$ and $n_{\p}$) and on $\alpha$,
\item We use a linear interpolation in the $\Delta^2$ dimension on
  $\Psi$, $n_{\n}$, $n_{\p}$ and $\alpha$.
\end{enumerate}

\begin{figure*}[h]
\center 
\includegraphics[width = 0.45\textwidth]{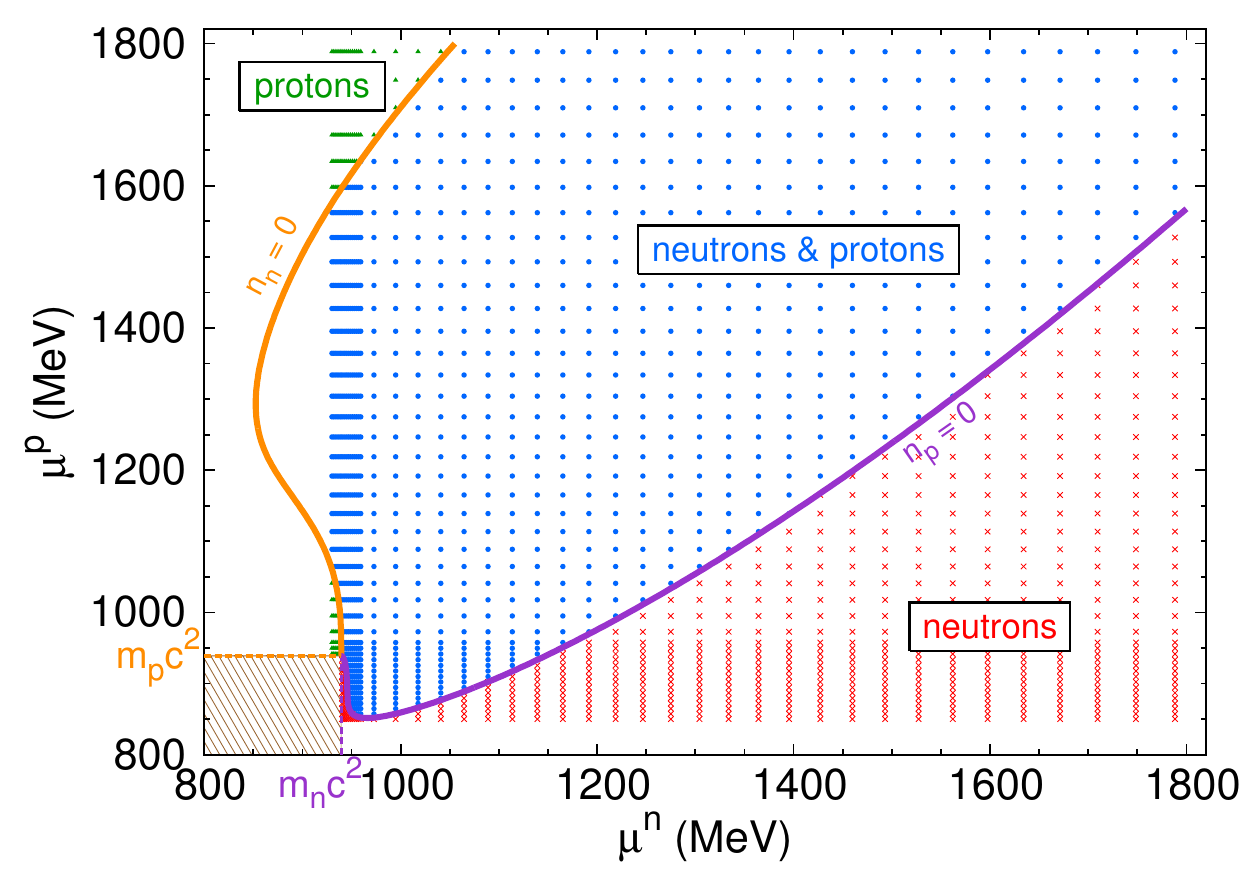}
\includegraphics[width = 0.45\textwidth]{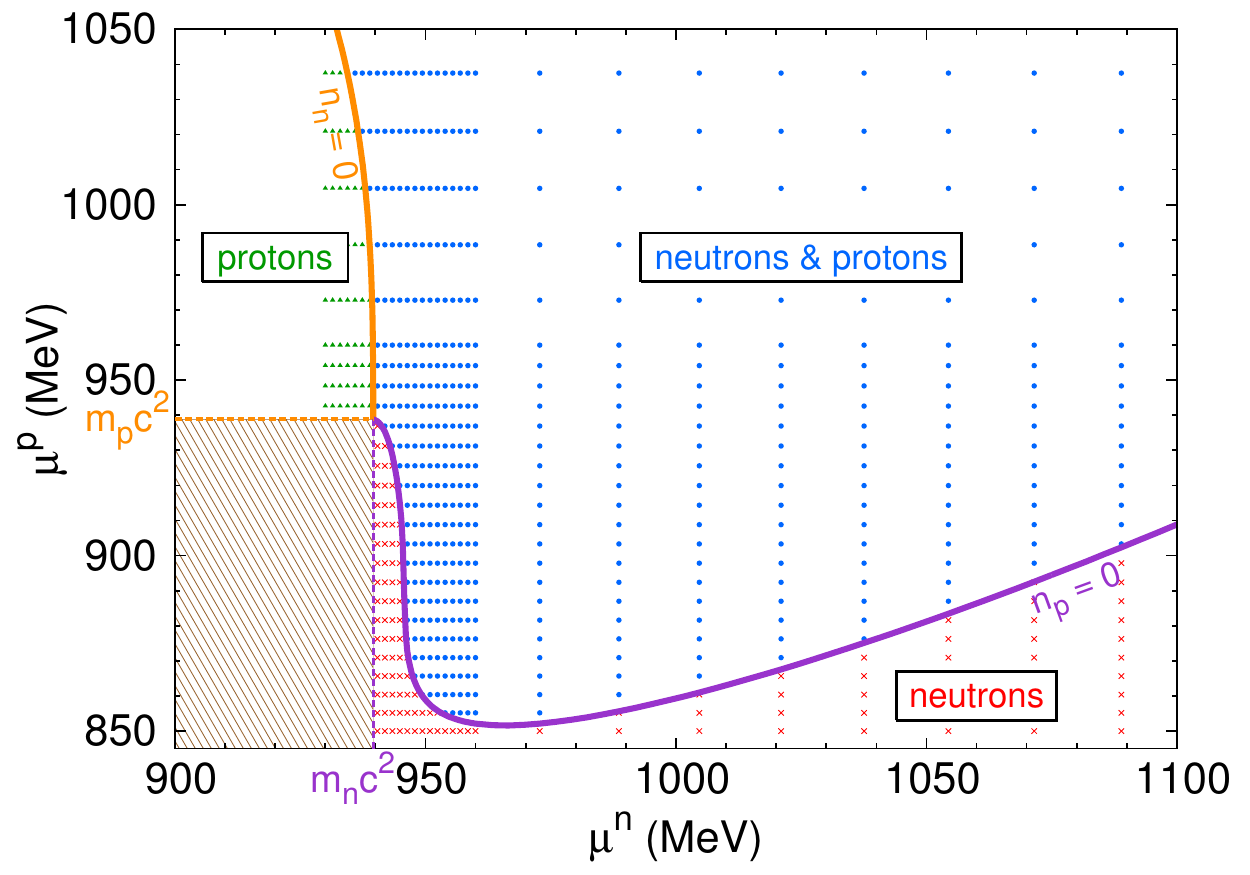}
\caption{Single-fluid (red crosses and green triangles) and two-fluid
  (blue dots) areas in the plane $(\mu^{\n}, \mu^{\p})$, for
  $\Delta^2= 0$, using the DDH$\delta$ EoS. The two-fluid zone is
  delimited by the limit lines $n_{\n}=0$ (orange) and $n_{\p}=0$
  (purple), beyond which one fluid disappears. For chemical potentials
  below the rest masses $m_{\n} = 939.6$ MeV and $m_{\p} = 938.8$ MeV
  no fluid is present, as can be seen on the zoom shown on the
  right. Note that $m_{\p}$ denotes here the sum of the rest mass of
  protons and that of electrons since the subscript $p$ stands here
  for fluid of charged particles. For better clarity, only a small
  proportion of the data contained in the table is plotted. In order
  to describe the area at low densities, where rapid variations occur,
  one uses a refined mesh. As realistic configurations are expected to
  be close to $\beta$-equilibrium and corotation, only the data around
  the $\mu^{\n} = \mu^{\p}$ line will be used.}
\label{EOS_cut}
\end{figure*}

To use the 2D interpolation method in \textcolor{blue}{\textbf{2.}},
it is necessary to provide some values of the function, its two
derivatives and the cross-derivative in the table. In the case of
$\alpha$, this cross-derivative would be a third-order derivative in
$\Psi$, that can not be provided with a good precision. Thus, for
simplicity, we employ the same interpolation scheme for $\Psi$ and
$\alpha$, without considering the cross-derivative in the second
case. The precision on the global interpolation scheme remains
sufficiently good. Note that we simply used a linear interpolation in
the relative speed because the data provided in the table are computed
with a first-order method. No derivatives with respect to $\Delta^2$
are thus required in the table.

We studied the consistency of this interpolation scheme by comparing
the results given by the code using directly an analytic EoS, as was
studied in \citep{prix2005relativistic}, and by the same code
interpolating a table based on the same EoS (computed with
machine-precision). The relative difference in the numerical results
obtained within these two methods were found to be very small.

A part of the DDH$\delta$ table, corresponding to the $\Delta^2=0$
plane, is shown in Fig.~\ref{EOS_cut}. The different areas in which
protons and/or neutrons are present are displayed. As can be seen in
Fig.~\ref{EOS_cut}, neutrons (and protons) can appear in the system
for values of the chemical potential below the corresponding rest
mass, as a consequence of the strong interactions between nucleons
(see Sec.~\ref{EOS}).

\bibliography{biblio}

\end{document}